\documentclass[prc,twocolumn,superscriptaddress,showpacs]{revtex4-1}
\usepackage{graphicx,amsmath,amssymb,bm,tabularx,tabulary,color}

\newcommand{\be}{\begin{equation}}	
\newcommand{\ee}{\end{equation}}
\newcommand{\vlowk}{V_{{\rm low}\,k}}
\newcommand{\lm}{\Lambda}
\newcommand{\fmi}{\, \text{fm}^{-1}}
\newcommand{\mev}{\, \text{MeV}}
\newcommand{\kev}{\, \text{keV}}
\newcommand{\la}{\langle}
\newcommand{\ra}{\rangle}
\newcommand{\hw}{\hbar\omega}
\newcommand{\pfg}{pfg_{9/2}}
\newcommand{\sdfp}{sdf_{7/2}p_{3/2}}

\newcommand{\fs}{f_{7/2}}
\newcommand{\pt}{p_{3/2}}
\newcommand{\gn}{g_{9/2}}
\newcommand{\ff}{f_{5/2}}
\newcommand{\po}{p_{1/2}}
\newcommand{\Q}{\widehat{Q}}
\newcommand{\Sb}{\widehat{S}}

\begin{document}

\title{Three-nucleon forces and spectroscopy of neutron-rich calcium isotopes}

\author{J.\ D.\ Holt}
\email{jason.holt@physik.tu-darmstadt.de}
\affiliation{Institut f\"ur Kernphysik, Technische Universit\"at
Darmstadt, 64289 Darmstadt, Germany}
\affiliation{ExtreMe Matter Institute EMMI, GSI Helmholtzzentrum f\"ur
Schwerionenforschung GmbH, 64291 Darmstadt, Germany}
\affiliation{National Superconducting Cyclotron Laboratory
and Department of Physics and Astronomy, Michigan State University,
East Lansing, MI 48844, USA}

\author{J.\ Men\'{e}ndez}
\email{javier.menendez@physik.tu-darmstadt.de}
\affiliation{Institut f\"ur Kernphysik, Technische Universit\"at
Darmstadt, 64289 Darmstadt, Germany}
\affiliation{ExtreMe Matter Institute EMMI, GSI Helmholtzzentrum f\"ur
Schwerionenforschung GmbH, 64291 Darmstadt, Germany}

\author{J.\ Simonis}
\email{simonis@theorie.ikp.physik.tu-darmstadt.de}
\affiliation{Institut f\"ur Kernphysik, Technische Universit\"at
Darmstadt, 64289 Darmstadt, Germany}
\affiliation{ExtreMe Matter Institute EMMI, GSI Helmholtzzentrum f\"ur
Schwerionenforschung GmbH, 64291 Darmstadt, Germany}

\author{A. Schwenk}
\email{schwenk@physik.tu-darmstadt.de}
\affiliation{Institut f\"ur Kernphysik, Technische Universit\"at
Darmstadt, 64289 Darmstadt, Germany}
\affiliation{ExtreMe Matter Institute EMMI, GSI Helmholtzzentrum f\"ur
Schwerionenforschung GmbH, 64291 Darmstadt, Germany}

\begin{abstract}

We study excited-state properties of neutron-rich calcium isotopes
based on chiral two- and three-nucleon interactions. We first discuss
the details of our many-body framework, investigate convergence
properties, and for two-nucleon interactions benchmark against
coupled-cluster calculations. We then focus on the spectroscopy
of $^{47-56}$Ca, finding that with both 3N forces and an extended
$\pfg$ valence space, we obtain a good level of agreement with
experiment. We also study electromagnetic transitions and find that
experimental data are well described by our calculations. In addition,
we provide predictions for unexplored properties of neutron-rich
calcium isotopes.

\end{abstract}


\maketitle

\section{Introduction}

Understanding the evolution of shell structure from the valley of
stability to neutron-rich extremes represents a key challenge in
nuclear structure~\cite{Baumann,Sorlin}. With a closed proton shell, the
calcium isotopes provide an ideal region to investigate shell formation
and evolution in medium-mass nuclei from nuclear
forces. The rich shell structure beyond $^{48}$Ca, combined with the
capabilities of rare isotope beam facilities, has lead to an intensive
focus on calcium isotopes, where new measurements provide exciting
tests and constraints for state-of-the-art many-body methods and
nuclear forces.

Recent experiments have established new shell closures in exotic
calcium isotopes. A possible $N=32$ closure in $^{52}$Ca was first
recognized from a higher first-excited $2^+$ ($2^+_1$)
energy~\cite{Huck,Gade}, more than $1.5\mev$ higher than the $2^+_1$
energy in the neighboring $^{50}$Ca. Moreover, signatures of a $N=32$
magic number were discovered in nearby titanium and chromium
isotopes~\cite{Janssens,Mantica,Crawford,Cr}. These observations have
been complemented by high-precision mass measurements, which revealed
a flat behavior of the two-neutron separation energy, $S_{2n}$,
leading up to the shell closure at $^{52}$Ca~\cite{Gallant}.
Groundbreaking mass measurements out to $^{54}$Ca~\cite{Wienholtz}
discovered a steep decrease in $S_{2n}$ from $^{52}$Ca to
$^{54}$Ca. Combined with the high two-neutron shell-gap
$S_{2n}(Z,N)-S_{2n}(Z,N+2)$, this unambiguously established $N=32$ as
a prominent shell closure~\cite{Wienholtz}. Evidence for a $N=34$
shell closure has proven more elusive, as the $2^+_1$ energy is not
high in the neighboring titanium or chromium
isotopes~\cite{Marginean,Liddick,Dinca,Rejmund}. Very recently, the
$2^+_1$ energy in $^{54}$Ca was found to be only $\sim 500\kev$ below
that in $^{52}$Ca~\cite{Steppenbeck}, suggesting a new shell closure.
Mass measurements through $^{56}$Ca will be essential to establish the
closed-shell nature of $^{54}$Ca. In addition, the spectroscopy of
neutron-rich calcium isotopes provides valuable information, with
important tests of theoretical calculations~\cite{Perrot,Fornal,Montanari}.

Previous work for the calcium isotopes includes phenomenological
valence-space~\cite{SMRMP,KB3G,GXPF1,GXPF1A,GXPF1B} or
beyond-mean-field calculations~\cite{Rodriguez}. While these
approaches are generally successful in reproducing experiment up to
$^{52}$Ca, disagreement is significant in $^{53,54}$Ca, which led to
many experimental and theoretical efforts aiming to clarify the nature
of $N=34$ in calcium. The uncertainty in extrapolating
phenomenological models to exotic nuclei shows the importance of
developing systematic many-body approaches based on nuclear
forces. Such calculations were initially pursued based on two-nucleon
(NN) forces, but failed to reproduce the standard $N=28$ shell closure
and other key features in calcium for $A \gtrsim
47$~\cite{Gmatrix,SMRMP}.  Neglected three-nucleon (3N) forces were
suggested to be the crucial missing ingredient~\cite{Zuker}.

The calcium region currently represents a frontier for ab initio
calculations based on NN and 3N forces~\cite{3NRMP}. The first
application of 3N forces in calcium was in the context of
valence-space Hamiltonians~\cite{Calcium}, which demonstrated the
important role of 3N forces in reproducing the dripline and spectra in
oxygen isotopes~\cite{Oxygen,Oxygen2}, as well as in proton-rich
nuclei~\cite{protonrich}. In calcium isotopes, 3N forces provided the
first microscopic explanation for the $N=28$ magic number in
$^{48}$Ca, as well as for the ground-state
energies~\cite{Calcium}. The improved calculations of
Refs.~\cite{Gallant,Wienholtz,pairing} successfully predicted the
additional binding found in $^{52}$Ca~\cite{Gallant} as well as the
behavior of the two-neutron separation energy from
$^{48-54}$Ca~\cite{Wienholtz}. Coupled-cluster (CC) calculations,
including continuum degrees of freedom and phenomenological 3N forces,
also found a very good description of these
signatures~\cite{CCCa}. This agreement extends to ab initio
self-consistent Green's function (SCGF) calculations with chiral NN
and 3N forces~\cite{GGF,GGF2}. In addition, the CC calculations of
Ref.~\cite{Roth} and the ab initio in-medium similarity
renormalization group (IMSRG)~\cite{IMSRG,Heiko} have been applied to
the calcium isotopes.

In this paper, we present a comprehensive study of excited-state
properties of neutron-rich calcium isotopes based on our valence-space
approach of Refs.~\cite{Oxygen,Oxygen2,Calcium,Gallant,Wienholtz,%
protonrich,pairing}, focusing in particular on shell structure in
the region around $N=28,32,34$. In Sect.~\ref{formalism}, we discuss
details of the calculation of valence-space Hamiltonians
perturbatively, based on NN and 3N forces from chiral effective field
theory (EFT)~\cite{EFTRMP,EMPR}. We investigate convergence both
order-by-order in many-body perturbation theory and in terms of
intermediate-state excitations, with NN interactions evolved to low
momentum via renormalization group (RG) methods. For NN interactions,
we benchmark against CC calculations and find reasonable agreement. In
Sect.~\ref{results}, we calculate spectra and electromagnetic
transitions in $^{47-56}$Ca, showing that with 3N forces and an
extended valence space, good agreement with experiment is obtained, in
many cases comparable to phenomenological interactions. Finally, we
explore the role of residual 3N forces. Similar to the oxygen
isotopes~\cite{Caesar}, their impact on spectra is minor, while for
ground-state energies their contributions increase with the number of
valence nucleons.

\section{Microscopic valence-space Hamiltonians}
\label{formalism}

\subsection{Many-body perturbation theory}

For a given nucleus, the solution of the $A$-body Schr\"{o}dinger
equation with the Hamiltonian $H$ gives the eigenstates $|\psi_n\ra$
and energies $E_n$,
\be
H|\psi_n\ra = (H_0+V)|\psi_n\ra=E_n|\psi_n\ra \,,
\ee
where $H_0$ defines a single-particle basis $|\phi_n\ra$ with 
corresponding eigenvalues $\epsilon_n$,
\be
H_0|\phi_n\ra = \epsilon_n|\phi_n\ra \,, 
\ee
and $V$ includes the interactions between nucleons. Solving the
$A$-body Schr\"{o}dinger equation in a large single-particle basis by
diagonalization is challenging due to the large number of
configurations involved. Therefore, many-body methods generally take
one of two strategies to describe medium-mass nuclei. In approaches
such as CC theory~\cite{CCCa,Roth,CCreview}, SCGF
theory~\cite{GGF,GGF2}, or the IMSRG~\cite{IMSRG,Heiko}, all nucleons
are active, but some truncations are necessary in practice. In
valence-space methods, the number of degrees of freedom is reduced by
treating the nucleus as a many-body system comprised of a closed-shell
core, with the additional nucleons occupying a truncated
single-particle (valence) space. After deriving an effective
valence-space Hamiltonian, this is then diagonalized exactly in the
valence space.

We first define operators $P$ and $Q$, which project into and out of
the valence space, respectively,
\begin{align}
P &= \sum_{i=1}^d |\phi_i\ra \la\phi_i| \,, \\
Q &= 1-P \,,
\end{align}
where $d$ is the dimension of the valence space, $P^2=P$, $Q^2=Q$, and 
$PQ=0$. Then, the goal is to construct an effective valence-space 
Hamiltonian, $H_{\rm eff}$, 
\be
PH_{\rm eff}P|\psi_\alpha\ra = E_\alpha P|\psi_\alpha\ra \,,
\ee
with
\be
H_{\rm eff} = \sum^d_{i=1} \varepsilon_i a^{\dagger} a + V_{\rm eff} \,,
\ee
which after diagonalization in the valence space reproduces a subset
$E_\alpha$ of eigenvalues of the full $A$-body Hamiltonian.  Here,
$\varepsilon_i$ are the single-particle energies (SPEs) of the $d$
orbitals in the valence space and $V_{\rm eff}$ is the effective
interaction between valence nucleons.

Many-body perturbation theory (MBPT) provides a diagrammatic framework
to calculate both the SPEs $\varepsilon_i$ and $V_{\rm eff}$ from
nuclear forces~\cite{Gmatrix,LNP,Scott,Coraggio}. This approach has
been pursued with NN interactions, but due to poor agreement with
experiment, all shell model calculations in practice involve
adjustments of either the SPEs, $V_{\rm eff}$, or both. To calculate
$H_{\rm eff}$ we start from an energy-dependent effective interaction
between valence nucleons, the $\Q$-box, which takes into account
excitations outside the valence space,
\be
\Q(\omega)=PVP+PVQ\frac{1}{\omega-QHQ}QVP \,,
\ee
and is evaluated at the unperturbed starting energy $\omega =
PH_0P$. The diagrammatic expansion of $\Q$ consists of all
irreducible, valence-linked diagrams. To remove the energy dependence,
we include folded diagrams through a nonperturbative transformation
involving $\Q$ and its energy derivatives. This results in an
energy-independent, size-extensive effective interaction
\be
V_{\rm eff}^{(k)}=\Q+\sum_{m=1}^{\infty}\frac{1}{m!}
\biggl(\frac{\text{d}^m\Q}{\text{d} \omega^m} \biggr)
\Bigl( V_{\rm eff}^{(k-1)} \Bigr)^m \,.
\ee
This integral equation is solved by iteration, which converges when
$V_{\rm eff}^{(k)} \approx V_{\rm eff}^{(k-1)}$, typically after $\sim
15$ iterations. We make two approximations in our evaluation of
$\Q$. First we truncate $\Q$ at some finite order. In this work we
include contributions up to third order in MBPT, the current
state-of-the art.  Second, excitations out of the valence space are
allowed to some finite energy $N \hw$, which is ultimately limited by
the size of the single-particle basis. Convergence with respect to
these two approximations is discussed in Sect.~\ref{convergence}.

\begin{figure*}
\begin{center}
\includegraphics[scale=0.65,clip=]{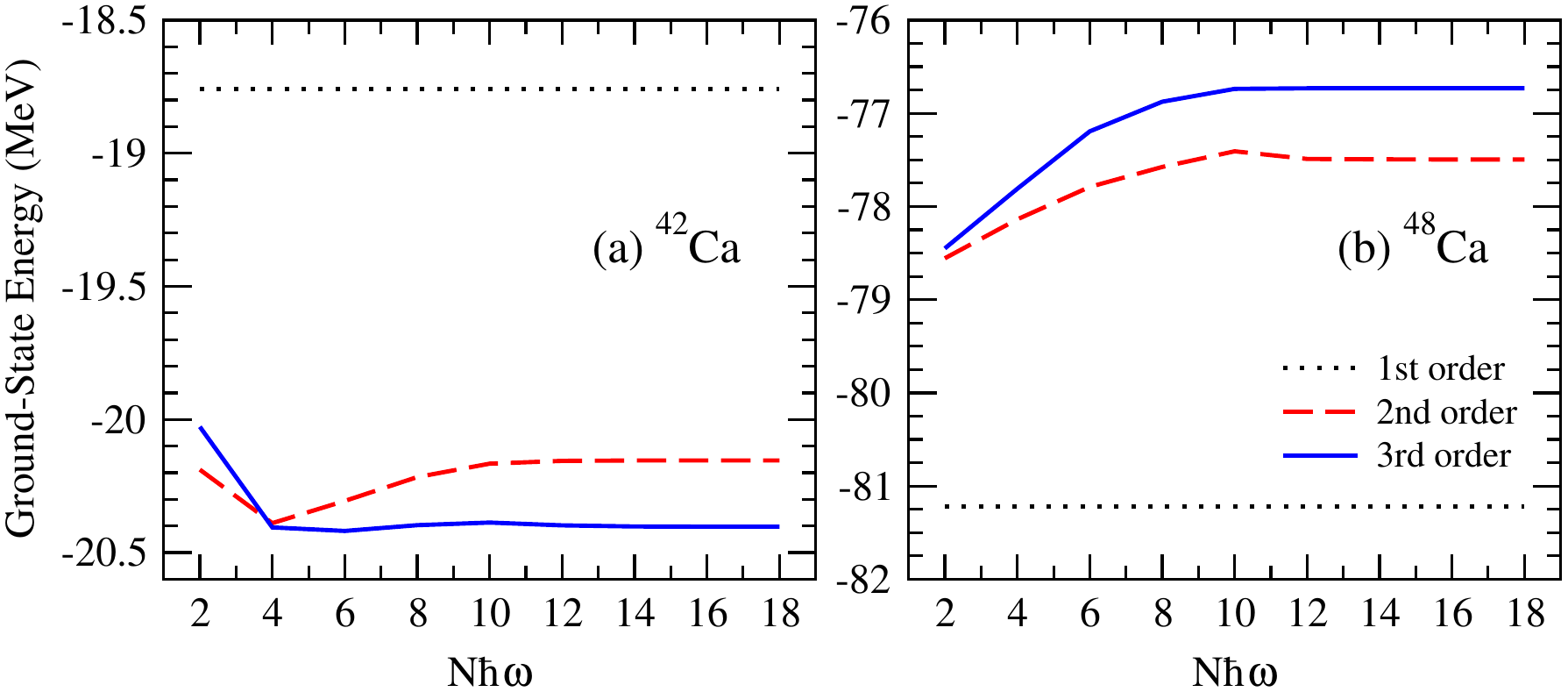}
\end{center}
\vspace{-2mm}
\caption{(Color online) Convergence of the (a) $^{42}$Ca and (b) $^{48}$Ca
ground-state energies as a 
function of increasing intermediate-state excitations $N \hw$ and
perturbative order. Calculations are based on NN forces in 13 major
harmonic-oscillator shells.\label{gs_conv}}
\end{figure*}

We then calculate the SPEs in $^{41}$Ca consistently from the
corresponding set of one-body diagrams, the $\Sb$-box, also taken to
third order in MBPT in the same harmonic-oscillator basis as the
$\Q$-box. To obtain the single-particle energies, we solve the coupled
Dyson equations,
\be
\varepsilon_i^{(k)} = \la i | H_0 | i \ra + \Sb_i(\varepsilon_i^{(k-1)}) \,,
\ee
by iteration starting from $\varepsilon_i^{(0)}=\omega$ until
$\varepsilon_i^{(k)} \approx \varepsilon_i^{(k-1)}$. Because the
$\varepsilon_i$ depend on relative shifts in the unperturbed
harmonic-oscillator spectrum, we also update the unperturbed
valence-space energy to be the centroid of the converged SPEs and
iterate until the centroid of the final SPEs is equal to the
unperturbed value. Convergence is reached after $\sim 10$ iterations.

\subsection{Nuclear interactions}

Our results are based on nuclear forces derived in chiral EFT, a
systematic expansion for nuclear forces~\cite{EFTRMP,EMPR} in which 3N
interactions arise naturally at next-to-next-to-leading order
(N$^2$LO). At the NN level, we perform an RG
evolution~\cite{Vlowk1,Vlowk2} of the $500\mev$ N$^3$LO NN potential
of Ref.~\cite{N3LO} using a smooth regulator~\cite{smooth} with $\lm =
2.0\fmi$, to obtain a low momentum interaction, $\vlowk$. The RG
evolution decouples low from high momenta and improves the convergence
of the MBPT calculation~\cite{Vlowk2}. At the 3N level, we use the
leading N$^2$LO 3N forces~\cite{chiral3N1,chiral3N2}, which include a
long-range two-pion-exchange part, a shorter-range one-pion exchange,
and a 3N contact interaction. The two undetermined 3N couplings are
fit to the $^3$H binding energy and the $^4$He radius at the same
resolution scale as the $\vlowk$ interaction~\cite{3Nfit}.

In the calculation of valence-space Hamiltonians, we include
normal-ordered one- and two-body parts of 3N forces, which correspond
to interactions among one valence and two core nucleons, or two
valence and one core nucleon, respectively.  These give rise to
repulsive interactions among valence neutrons and increase the
spin-orbit splitting in the SPEs~\cite{Oxygen,Calcium}. They are
expected to be dominant over residual 3N forces between three valence
particles because of phase-space considerations~\cite{Fermi}. This has
been confirmed in CC calculations of light and medium-mass
nuclei~\cite{CC3N,Roth}. In Sect.~\ref{residual3N}, we explore this by
calculating the contributions from residual 3N
forces~\cite{Caesar,Wienholtz} for ground and first-excited states.

We work in a harmonic-oscillator basis with $\hw=11.48\mev$ and
include NN forces in 13 major shells and 3N forces in 5 major
shells. When 3N forces are included fully to third order in MBPT, we
find that the contribution to SPEs range from $\sim 1-6 \mev$ for
different orbitals, approximately an order of magnitude larger than
the effects on valence-space interactions, as expected from the
hierarchy of normal-ordered contributions~\cite{CC3N}. Moreover, with
3N forces, the calculated SPEs are comparable to the empirical values
of the phenomenological GXPF1A~\cite{GXPF1A} and KB3G~\cite{KB3G}
interactions for the $pf$-shell orbitals~\cite{Calcium,pairing}, as
shown in Table~\ref{spetab}.

\begin{table}
\begin{center}
\begin{tabular*}{0.98\columnwidth}{@{\extracolsep{\fill}}cccc}
\hline\hline
Orbital & \multicolumn{2}{c}{Phenomenological} & MBPT \\
& GXPF1A~\cite{GXPF1A} & KB3G~\cite{KB3G} & $\rm pfg_{9/2}$ \\
\hline
$\rm f_{7/2}$ & $-8.62$ & $-8.60$ & $-8.05$ \\
$\rm p_{3/2}$ & $-5.68$ & $-6.60$ & $-5.86$ \\
$\rm p_{1/2}$ & $-4.14$ & $-4.60$ & $-3.22$ \\
$\rm f_{5/2}$ & $-1.38$ & $-2.10$ & $-1.33$ \\
$\rm g_{9/2}$ & $(-1.00)$ & -- & $-1.23$ \\
\hline\hline
\end{tabular*} 
\end{center}
\vspace{-2mm}
\caption{Phenomenological and calculated (MBPT) SPEs in MeV. 
Details are given in the text.\label{spetab}}
\end{table}

\begin{figure*}
\begin{center}
\includegraphics[scale=0.65,clip=]{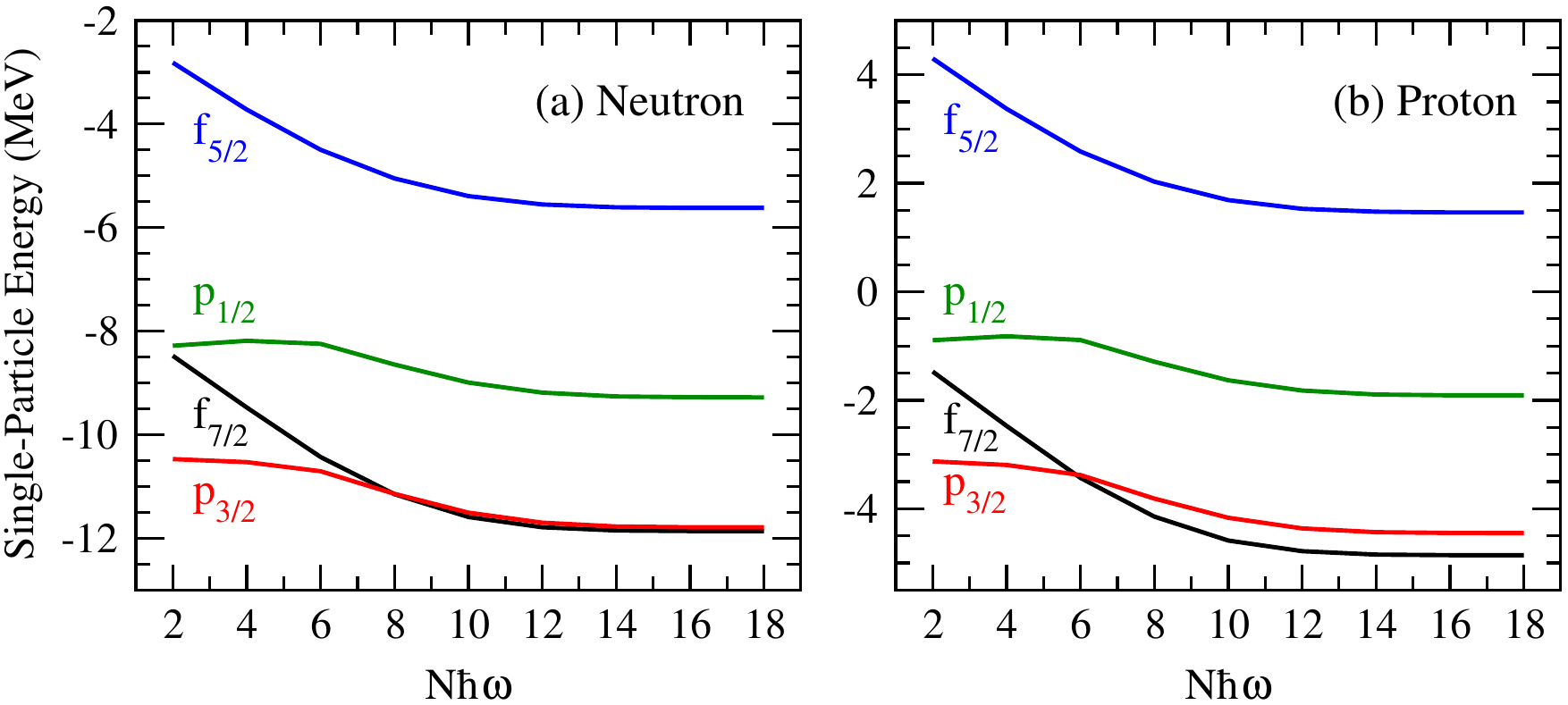}
\end{center}
\vspace{-2mm}
\caption{(Color online) Convergence of (a) neutron and (b) proton SPEs
as a function of increasing intermediate-state excitation $N \hw$.
Calculations are based on NN forces in 13 major harmonic-oscillator
shells.\label{spe_conv}}
\end{figure*}

\subsection{Convergence properties}
\label{convergence}

Next, we discuss the convergence properties of MBPT when using
low-momentum NN interactions. For a fixed valence space, the results
must converge order-by-order in the $\Q$- and $\Sb$-box expansion as
well as in the allowed intermediate-state excitations. Because the RG
evolution renders nuclear interactions more perturbative and decouples
low- from high-oscillator states, improved convergence behavior is
expected in both respects~\cite{Vlowk2}.

Assessing order-by-order convergence beyond third order is a challenge
for MBPT, since complete fourth-order calculations are beyond current
computational capabilities and have not been attempted. In
Fig.~\ref{gs_conv}, we show the order-by-order convergence of the
$^{42,48}$Ca ground-state energies of $^{42,48}$Ca as a function of
increasing intermediate-state excitations $N\hw$, within a
harmonic-oscillator basis of 13 major shells. $N\hw$ denotes the
number of excitation quanta in a given intermediate-state
configuration (e.g., two neutrons excited three shells above the
valence space is a $6\hw$ excitation). In these studies, we use
NN-only forces with empirical SPEs and see promising order-by-order
behavior: the change from second to third order is $\sim 15\%$ of the
change from first to second order. While the calculations cannot be
said to be completely converged at third order, this trend suggests
that changes due to a complete fourth-order calculation would be
small. In particular, they will be less important than other
uncertainties in the calculation, such as the uncertainties in the
input Hamiltonian.

We also observe in Fig.~\ref{gs_conv} that in terms of
intermediate-state excitations, the convergence with $\vlowk$ is
rapid, with 3rd order converging faster than 2nd order. For all
orders, the ground-state energies of both $^{42,48}$Ca are well
converged by $\sim12\hw$. Similarly, Fig.~\ref{spe_conv} shows the
convergence of neutron and proton SPEs in the $pf$-shell as a function
of $N\hw$. While convergence is slower compared to the ground-state
energies, all SPEs are converged by $14\hw$, with neutron and proton
SPEs following a very similar convergence pattern. Finally, all
calculations with 3N forces seem to be converged when included in 5
major shells. In two-body matrix elements and SPEs, the
change from four to five major shells is $\sim 10\kev$ and $\sim
50\kev$, respectively.  Work to extend 3N forces 
beyond 5 major shells is currently in progress.

\subsection{Benchmark with coupled-cluster theory}

\begin{figure}
\begin{center}
\includegraphics[scale=0.68,clip=]{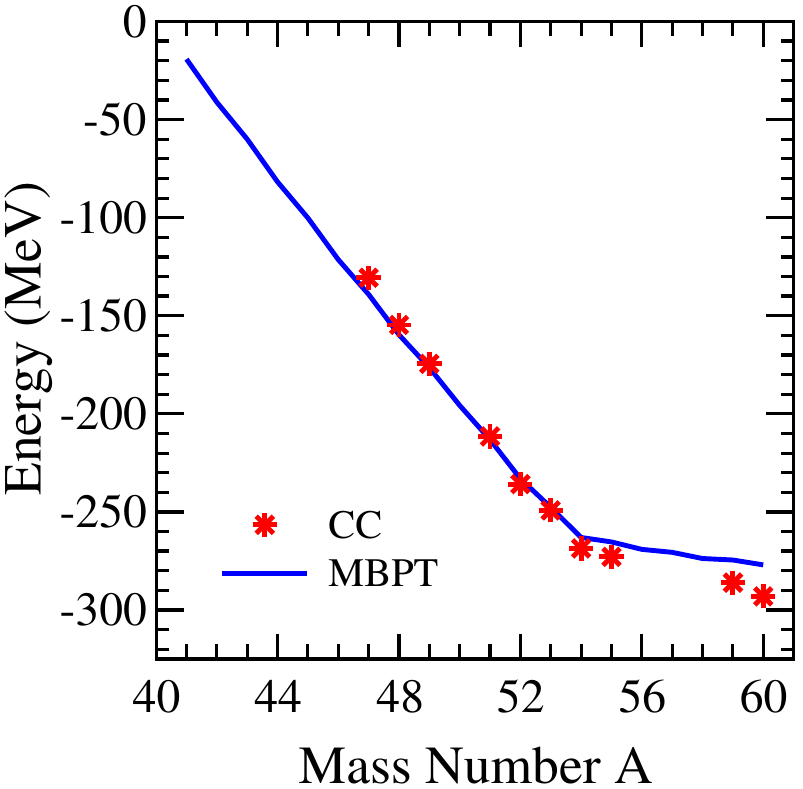}
\end{center}
\vspace{-2mm}
\caption{(Color online) Comparison of MBPT and CC ground-state energies of calcium
isotopes relative to $^{40}$Ca based on the same NN interaction (for
details see text). The MBPT results use the SPEs obtained in CC theory.
\label{CC}}
\end{figure}

We can also benchmark the MBPT energies with ab initio methods using
identical starting interactions and working in the same
single-particle basis. Here, we perform CC calculations for the
ground-state energies of the calcium isotopes using the same $\vlowk$
interaction in a single-particle basis of 13 major harmonic-oscillator
shells with $\hw = 12 \mev$. The results are shown in Fig.~\ref{CC}
relative to the ground-state energy of $^{40}$Ca. The closed
$j$-subshell systems, $^{40,48,52,54,60}$Ca, are calculated at the
$\Lambda$-CCSD(T) level~\cite{CClong,CCreview}. The $A \pm 1$ systems,
$^{47,49,51,53,55,59}$Ca, are obtained with the CC
particle-attached/removed equations of motion method at the singles
and doubles level (PA/PR-EOM-CCSD)~\cite{CClong,CCreview}.

To compare with CC results, we perform the MBPT calculations in the
$pf$ shell, where the SPEs are taken as the PA-EOM-CCSD ($\fs$, $\pt$,
$\po$, $\ff$) energies in $^{41}$Ca. The particle-attached $\gn$ is
not of single-particle character, so the MBPT $pf$-shell comparison
provides the cleanest benchmark.
This comparison probes the two-body part of the valence-space
Hamiltonian, assessing the reliability of the convergence trend
illustrated in Fig.~\ref{gs_conv}.

In Fig.~\ref{CC}, we find that the
MBPT ground-state energies are within $5\%$ (in most cases much
better) of those of CC theory. This shows that MBPT can be comparable
to CC theory for $\vlowk$ interactions, provided that consistent SPEs
are employed.

While the CC ground-state energies agree well with MBPT to $^{55}$Ca,
this agreement deteriorates for heavier isotopes. The reason is that the CC
calculations begin to fill the $\gn$ orbit, which is lower in energy
than the calculated $\ff$. This makes a comparison of the CC and
$pf$-shell valence-space calculations unreliable for
$^{59,60}$Ca. Moreover, a benchmark in the $\pfg$ space is not
possible, because, as mentioned, the CC one-particle attached 
$\gn$ state in $^{41}$Ca is not of single-particle character.

\subsection{Valence-space calculations}

For neutron-rich oxygen and calcium isotopes, we have shown that it is
necessary in MBPT calculations of valence-space Hamiltonians to expand
the valence space beyond the standard one major harmonic-oscillator
shell~\cite{Oxygen2,protonrich,Calcium,pairing}. This takes into
account the effects of the additional orbitals nonperturbatively, so
that the general strategy is to make the valence space for
diagonalization as large as possible and include the contributions
beyond the valence space in MBPT, which converges better for larger
valence spaces.

In this work, we perform calculations in both the $0f_{7/2}$,
$1p_{3/2}$, $0f_{5/2}$, $1p_{1/2}$ valence space ($pf$ shell) and the
extended space including the $0g_{9/2}$ orbit ($\pfg$ valence space),
in both cases on top of a $^{40}$Ca core.  We take two approaches with
respect to SPEs: in all $pf$-shell calculations we use the empirical
GXPF1A SPEs, while for the $\pfg$ space we either use the GXPF1A values
(setting $\gn=-1.0\mev$), or the MBPT SPEs calculated consistently,
as shown in Table~\ref{spetab}. The shell model codes
ANTOINE~\cite{SMRMP,Antoine} and NATHAN~\cite{SMRMP} have been used
throughout this work.

The $\pfg$ space consists of orbitals beyond one major
harmonic-oscillator shell, which means that the center-of-mass (cm)
motion of the valence nucleons will not factorize in
general. Following Refs.~\cite{ZukerCM,Ca40CM}, we have investigated
possible center-of-mass (cm) contamination in our calculations by
adding a cm Hamiltonian, $\beta H_{cm}$, with $\beta=0.5$, to our
original Hamiltonian. This has a modest impact on excitation spectra,
where states can be affected up to $\sim 200\kev$. This difference
can be understood because the non-zero cm two-body matrix elements are
also relevant matrix elements of the MBPT calculation, and a clear
separation between these two effects is difficult. Similarly, we find
non-negligible $\langle H_{cm} \rangle$ values, which point to
possible cm admixture and/or non-negligible occupancies of the $\gn$
orbital.

\begin{figure*}
\begin{center}
\includegraphics[width=0.68\textwidth,clip=]{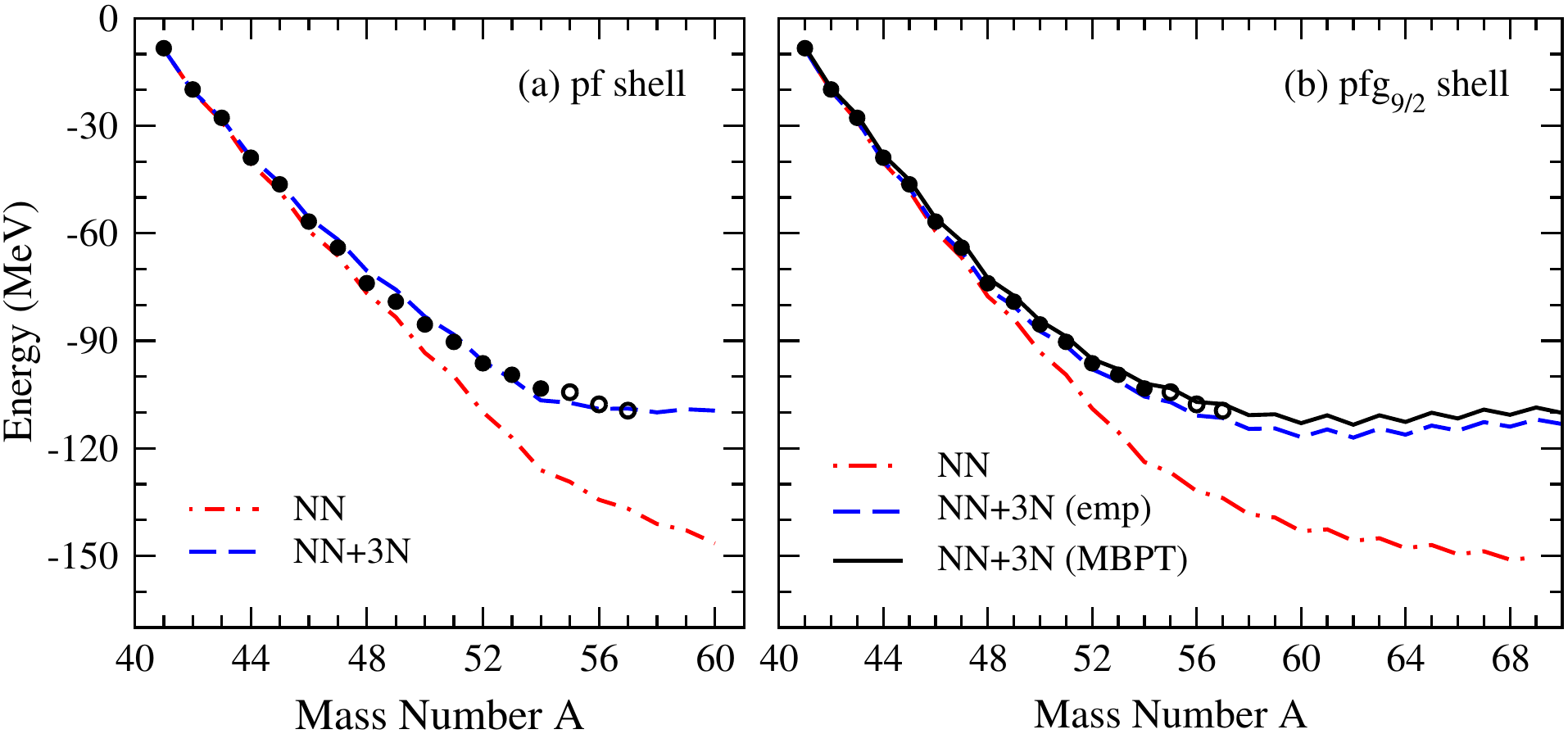}
\end{center}
\vspace{-2mm}
\caption{(Color online) Calculated ground-state energies of calcium isotopes in (a) $pf$ shell and
(b) $\pfg$ shell, compared
with experimental data (solid points) and AME2012 extrapolated values
(open circles) \cite{AME2012}. Calculations are performed in the extended $\pfg$
valence space and based on NN forces only, NN+3N forces with empirical
SPEs, and NN+3N forces with calculated (MBPT) SPEs.\label{gs}}
\end{figure*}

There are several directions in progress to investigate this further
in both the $\pfg$ and $\sdfp$~\cite{Oxygen2} spaces. We will carry
out a nonperturbative Okubo-Lee-Suzuki-Okamoto
transformation~\cite{Okubo,LS} into the standard one-major-shell
space, which is free of cm spurious states. This will keep the
treatment of the orbitals within the extended space nonperturbative,
while treating the MBPT configurations perturbatively. We will also
apply the IMSRG~\cite{IMSRG} to extended valence spaces, tailoring the
evolution so that the cross-shell matrix elements have small values,
$\langle H_{cm} \rangle \to 0$. Finally, we will explore different
valence spaces, choosing the core of the calculations so that the cm
factorizes. For instance, for the neutron-rich calcium isotopes a
$^{48}$Ca core can be used. Here, we follow the calculations of
ground-state energies of Refs.~\cite{Gallant,Wienholtz,pairing} and
present results for the spectra for the same interactions.

\section{Results}
\label{results}

\subsection{Ground-state energies}

The calculated ground-state energies for calcium isotopes are shown
for both the $pf$ and $\pfg$ shells in
Fig.~\ref{gs}. These are the same as for the predictions of the
neutron-rich $^{51-54}$Ca reported in
Ref.~\cite{Gallant,Wienholtz}. They update the results of
Ref.~\cite{Calcium}, where 3N forces where included only to first
order in MBPT. The repulsive effect of normal-ordered 3N
forces~\cite{Oxygen,Calcium} is evident in both valence spaces, 
and there is only a small difference between the calculations with empirical and calculated
(MBPT) SPEs, which reflects the similar values shown in
Table~\ref{spetab}.  

\begin{figure}
\begin{center}
\includegraphics[width=0.45\textwidth,clip=]{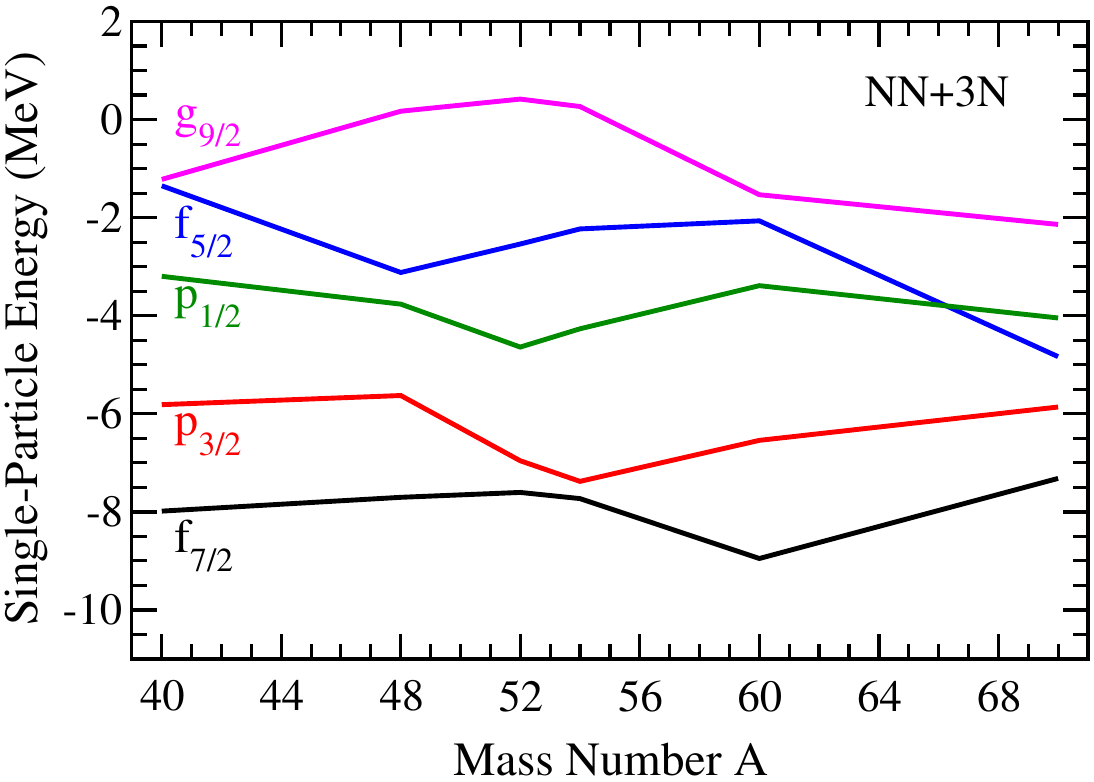}
\end{center}
\vspace{-2mm}
\caption{(Color online) Evolution of SPEs as a function of mass number. Calculations
are based on NN+3N forces in the extended $\pfg$ space.\label{espe}}
\end{figure}

While the $pf$ and $\pfg$ spaces give similar absolute ground-state 
energies, detailed comparisons to recent experimental two-neutron 
separation energies \cite{Gallant,Wienholtz} and three-point mass differences 
\cite{Gallant,pairing}
highlight the good agreement found with the $\pfg$-shell results.
Beyond
$^{60}$Ca the ground-state energies evolve very flat with $A$, which
makes a precise prediction of the dripline difficult. Moreover, for
masses beyond $^{54}$Ca, CC calculations indicate that continuum
degrees of freedom play an important role in lowering the $1d_{5/2}$
and $2s_{1/2}$ orbitals, not included in our calculations.  As a
result these orbitals may become degenerate with $0g_{9/2}$ near
$^{60}$Ca, and further lowering of the ground-state energies beyond
$^{60}$Ca is expected~\cite{CCCa}. Therefore, to explore reliably the
neutron-rich region towards the dripline, continuum degrees of freedom
and larger valence spaces are necessary.

\subsection{Spectra}
\label{spectra}

We now calculate the spectra of neutron-rich calcium isotopes,
comparing our MBPT predictions to experiment when available, as well
as to shell model results using the phenomenological interactions
GXPF1A~\cite{GXPF1A} and KB3G~\cite{KB3G}. We discuss in detail the
spectra of the neutron-rich isotopes $^{47-56}$Ca. For $^{47-50}$Ca we
present different calculations based on chiral interactions and
compare to experiment, emphasizing the importance of extended valence
spaces and 3N forces.  To quantify these effects in excitation
spectra, we show results with NN forces only and NN+3N forces, in the
$pf$-shell and in the extended $\pfg$ space. For $^{51-56}$Ca, where
there is none or limited experimental information on excited states, we focus
on the predictions of our best calculations (NN+3N forces in the
$\pfg$ space). The spectra for the lighter $^{42-46}$Ca, which mostly
probe the $f_{7/2}$ orbital, are given in Appendix~\ref{light_spectra}.

To understand these results, we refer to the effective single-particle
energies (ESPEs), given in Fig.~\ref{espe}, for NN+3N forces in the
$\pfg$ space, which describe the evolution of the spherical mean field
of the calculation. While correlations are important in the final
results and are included via exact diagonalization, ESPEs provide a
guide to the position of different orbitals for a given neutron number
within our valence-space framework.
The important role of the extended valence spaces is manifested in
the occupancies of the $\gn$ orbital. For ground-states, these become
significant in $^{47}$Ca, with $\gn$ occupation number 1.2, increasing to
2.1 in $^{56}$Ca. They are accompanied by a depletion of the $\fs$
occupations.

\subsubsection{$^{47}$Ca}

\begin{figure}
\begin{center}
\includegraphics[width=0.95\columnwidth,clip=]{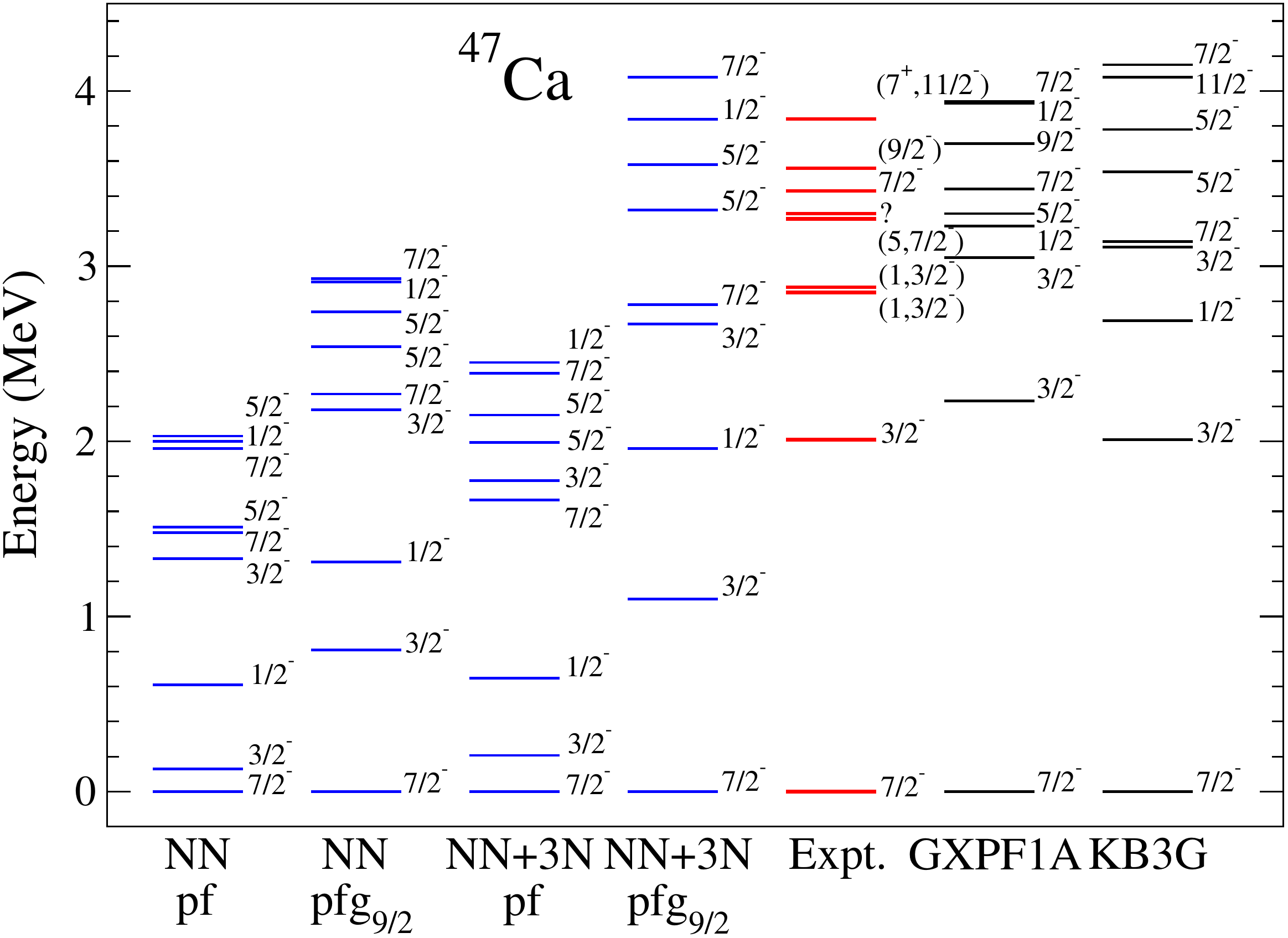}
\end{center}
\vspace{-2mm}
\caption{(Color online) Excitation energies of bound excited states in $^{47}$Ca 
compared with experiment~\cite{ENSDF} and phenomenological GXPF1A~\cite{GXPF1A}
and KB3G~\cite{KB3G} interactions. The NN-only results are calculated
in the $pf$ and $\pfg$ spaces with empirical SPEs. The NN+3N results
are obtained in the same spaces. In the $pf$-shell, empirical SPEs 
are used, while the $\pfg$ space results use the consistently 
calculated MBPT SPEs.\label{47}}
\end{figure}

In Fig.~\ref{47} we show the calculated spectra for $^{47}$Ca in the
$pf$ and $\pfg$ spaces using NN-only and NN+3N forces. In the
$pf$-shell calculations, the spectra are too compressed. The two
lowest-lying states differ by only $200 \kev$, and there is otherwise
very poor agreement with experiment.  Furthermore, the effects of 3N
forces in the $pf$-shell are relatively small.  Extending the
calculation to the $\pfg$ space with NN forces only partially improves
the spectrum, but it remains too compressed. Our final results with
NN+3N forces in the extended space improve the spectrum, leading to the
best agreement with experiment.

Nevertheless, we still observe deficiencies in our MBPT spectrum. The
major disagreement is in the lowest $3/2^-$ state, which is
approximately $1 \mev$ below experiment, reflecting the small
$f_{7/2}-p_{3/2}$ gap around $^{48}$Ca, as seen in
Fig.~\ref{espe}. This state is well reproduced by the phenomenological
interactions.  Similarly, the $1/2^-_1$ state is also low, due to the
small $f_{7/2}-p_{1/2}$ gap in our calculations.

\subsubsection{$^{48}$Ca}

\begin{figure}
\begin{center}
\includegraphics[width=0.95\columnwidth,clip=]{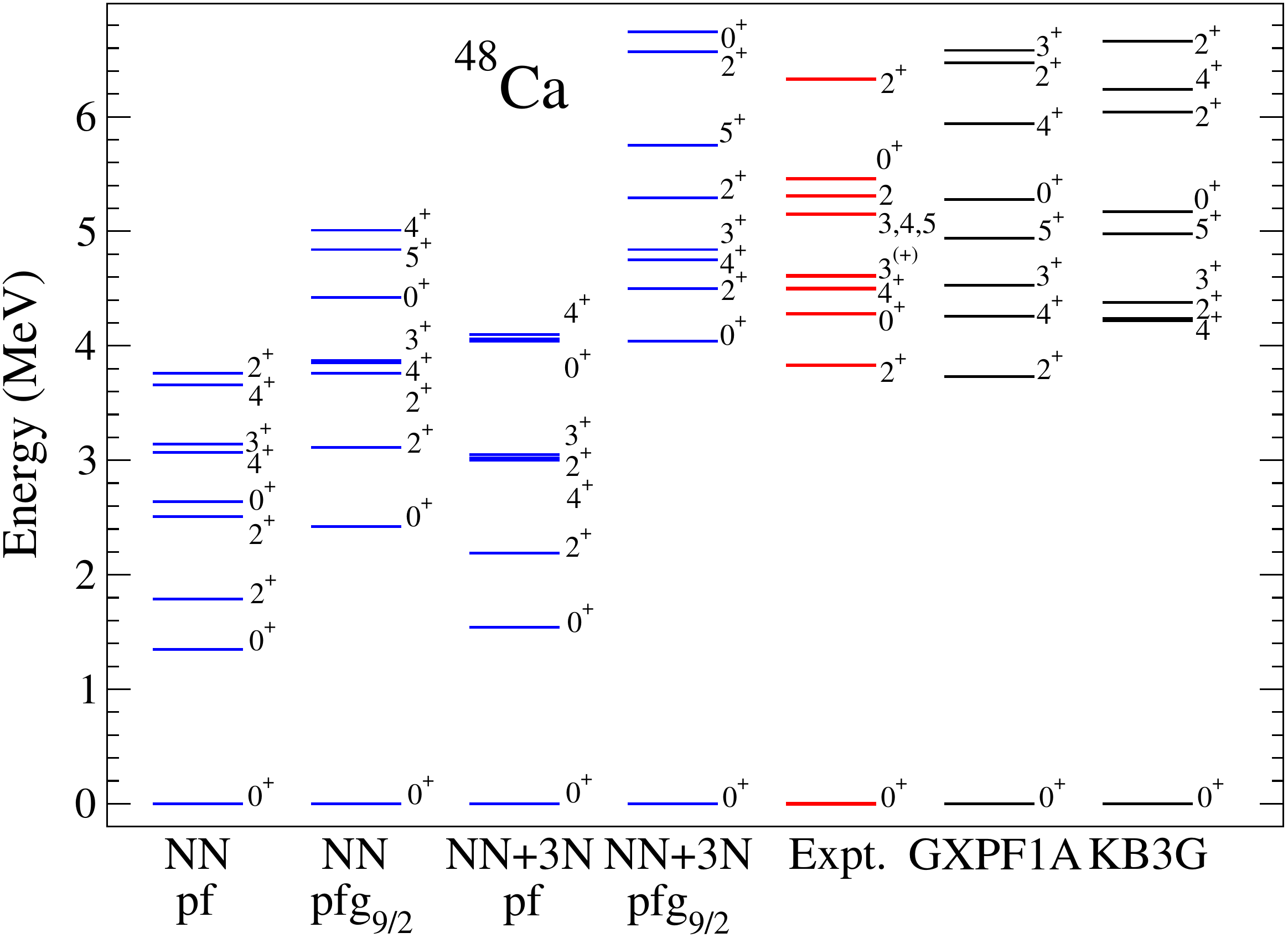}
\end{center}
\vspace{-2mm}
\caption{(Color online) Excitation energies of bound excited states in $^{48}$Ca 
compared with experiment~\cite{ENSDF} and phenomenological interactions 
(labels as in Fig.~\ref{47}).\label{48}}
\end{figure}

Figure~\ref{48} shows the calculated $^{48}$Ca spectra compared with
experiment.  As with $^{47}$Ca we note that the $pf$ space generally
gives too compressed spectra, and 3N forces give only minor
improvements. In the extended $\pfg$ space, while NN forces also give
a poor experimental description, significant improvement is obtained
when 3N forces are included.

The gap between the ground state and the $2^+_1$ state, a measure of
the shell closure at $N=28$, is well reproduced, though somewhat
overpredicted by $500\kev$. As we have a relatively small
$f_{7/2}-p_{3/2}$ gap in the ESPEs in Fig.~\ref{espe}, the high
$2^+_1$ state is a result of correlations involving $\gn$, in
particular $f_{7/2}-g_{9/2}$. On the other hand, we find a $0^+$ as
the first excited state, contrary to experiment, in all
calculations. Since this state is dominated by the $2p-2h$
configuration with respect to the ground state of the form
$(f_{7/2})^{-2}(p_{3/2})^2$, this may be related to a too strong
$f_{7/2}-p_{3/2}$ pairing interaction. Other excited states are in
good agreement with experiment, and comparable to the results of
GXPF1A and KB3G.

\subsubsection{$^{49}$Ca}

\begin{figure}
\begin{center}
\includegraphics[width=0.95\columnwidth,clip=]{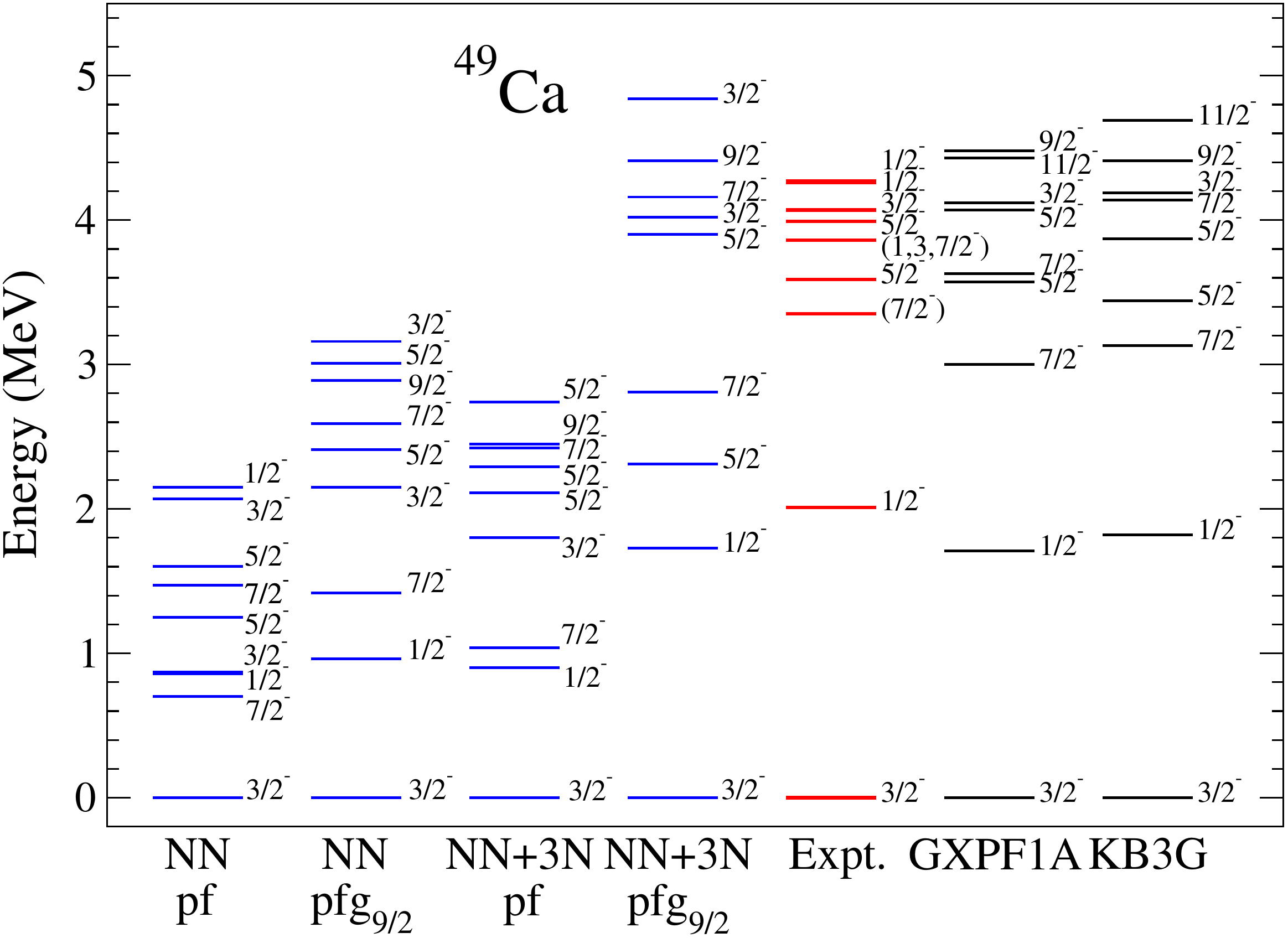}
\end{center}
\vspace{-2mm}
\caption{(Color online) Excitation energies of bound excited states in $^{49}$Ca
compared with experiment~\cite{Montanari,ENSDF} and phenomenological 
interactions (labels as in Fig.~\ref{47}).\label{49}}
\end{figure}

As in lighter isotopes, our calculations of $^{49}$Ca in Fig.~\ref{49}
show that with either NN forces only, or in the $pf$ shell, the
physics necessary to reproduce the spectrum is not adequately
captured; the excited states are too compressed and with incorrect
ordering. It is only in the NN+3N calculations in the extended $\pfg$
space that we observe a reasonable description of the $^{49}$Ca spectrum.

The ground state in $^{49}$Ca is dominated by the single-particle
configuration of a $p_{3/2}$ particle on top of $^{48}$Ca. Therefore,
the first excited $1/2^-_1$ state, predicted in very good agreement with
experiment, reflects the effective $p_{3/2}-p_{1/2}$ gap for this
nucleus. Also the location of the lowest $7/2^-_1$ state is in
reasonable agreement with the tentatively assigned experimental level
(it lies some $500\kev$ lower), and with predictions from the
phenomenological interactions.  This state is dominated by a $2p-1h$
$(f_{7/2})^{-1}(p_{3/2})^2$ configuration on top of $^{48}$Ca, and
therefore reflects the effective $f_{7/2}-p_{3/2}$ gap plus
correlations discussed for the closure of $^{48}$Ca.

However, in our calculations we observe that the $5/2^-_1$ state is
quite low compared to experiment and the phenomenological
interactions. This is indicative of a small effective
$p_{3/2}-f_{5/2}$ gap in this region. We also note that the spin of
the fourth excited state has not been experimentally identified, but
that our calculations, as in phenomenology, predict it as a $7/2^-$ state.

\subsubsection{$^{50}$Ca}

\begin{figure}
\begin{center}
\includegraphics[width=0.95\columnwidth,clip=]{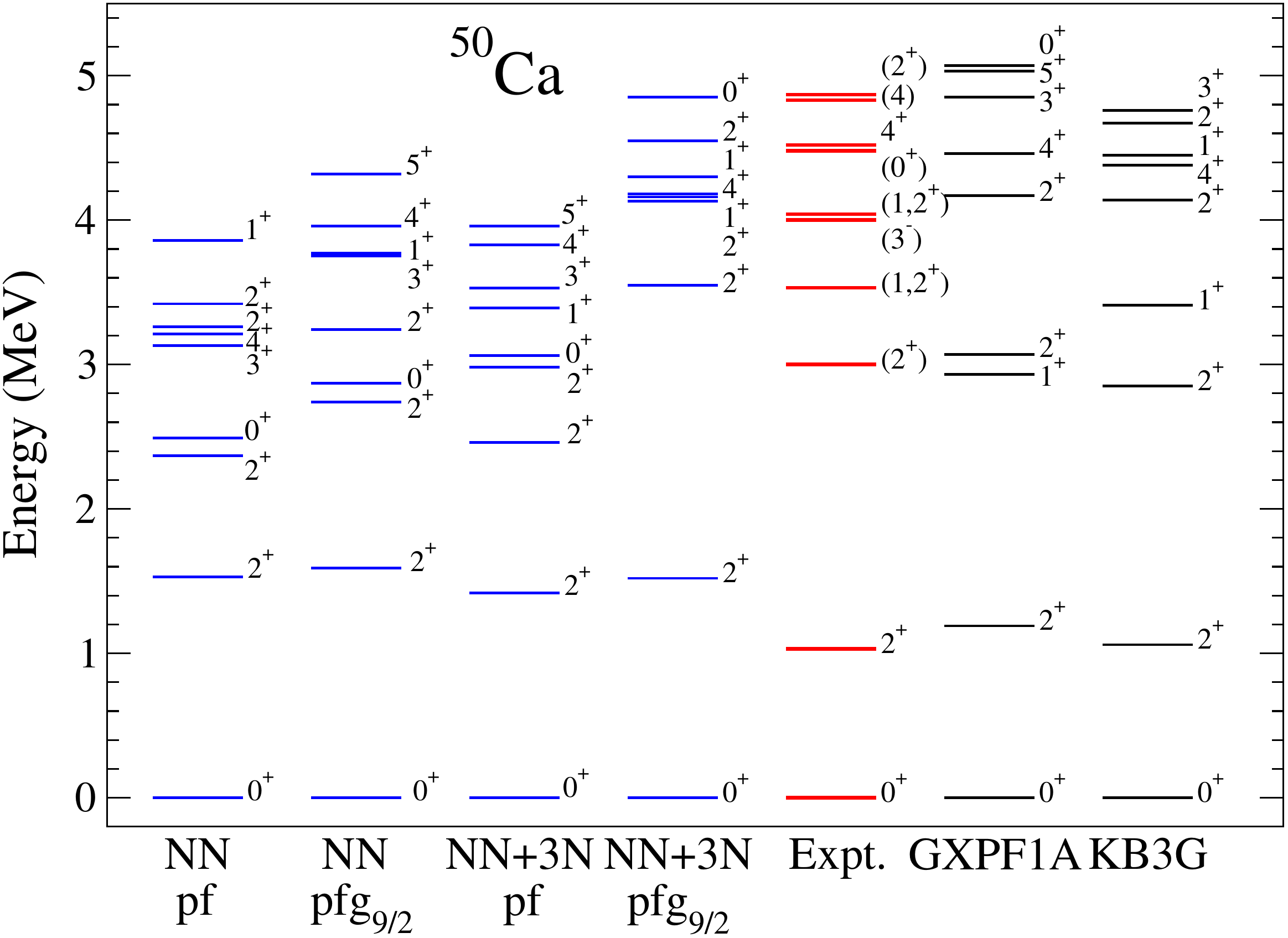}
\end{center}
\vspace{-2mm}
\caption{(Color online) Excitation energies of bound excited states in $^{50}$Ca
compared with experiment~\cite{Montanari,ENSDF} and phenomenological
interactions (labels as in Fig.~\ref{47}).\label{50}}
\end{figure}

In Fig.~\ref{50} we see that for $^{50}$Ca the location of the first
excited $2^+_1$ state is overpredicted in all MBPT calculations by
$\sim 500\kev$. The $0^+$ ground state and the $2^+_1$ state are
dominated by $(p_{3/2})^2$ configurations. Therefore, the increased
$2^+_1$ energy is related to the low excited $0^+$ state found in $^{48}$Ca.

Although most of the experimental spin and parity assignments are
tentative, in our calculations with NN+3N forces in the $\pfg$ space,
the remaining states are compatible with experiment and comparable to
the results with the phenomenological interactions.  In particular the
large $2\mev$ gap between the $2^+_1$ and $2^+_2$ states is not
reproduced in our other MBPT calculations. The location of the lowest
$1^+_1$ state differs significantly in the three calculations, which
are otherwise consistent with the data, with the MBPT prediction being
$1\mev$ and $500\kev$ above the GXPF1A and KB3G predictions,
respectively. A reliable assignment of the spin of the third excited
state in $^{50}$Ca at $3.53\mev$ is needed to identify this state and
test the theoretical calculations.

\subsubsection{$^{51}$Ca}

\begin{figure}
\begin{center}
\includegraphics[width=0.95\columnwidth,clip=]{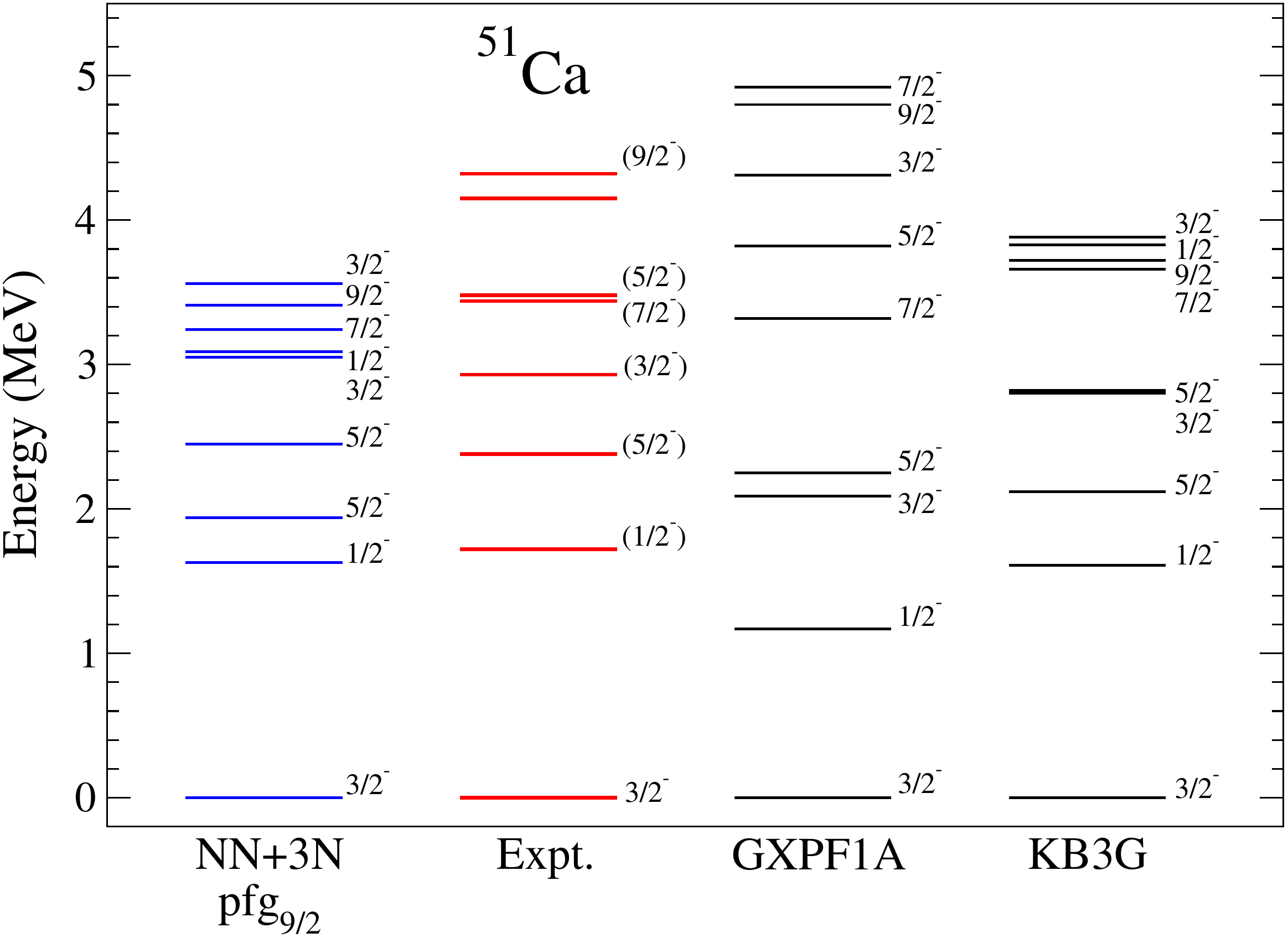}
\end{center}
\vspace{-2mm}
\caption{(Color online) Excitation energies of bound excited states in $^{51}$Ca
compared with experiment~\cite{Fornal,Rejmund} and phenomenological 
interactions (labels as in Fig.~\ref{47}).\label{51}}
\end{figure}

In $^{51}$Ca there is no definite experimental information on the
spins of the excited states, only tentative assignments based largely
on inferences from phenomenological
interactions~\cite{Rejmund,Fornal}. Therefore, we show in
Fig.~\ref{51} only our NN+3N calculation in the extended $\pfg$ space
and compare with the experimental excitation energies and with
the phenomenological results.

The $3/2^-$ ground state is dominated by a $p_{3/2}$ hole
configuration below the $N=32$ subshell closure. The first excited
$1/2^-$ state is indicative of the effective $p_{3/2}-p_{1/2}$ gap
(and approximate strength of the $N=32$ closure) and is in very good
agreement with the experimental tentative spin assignment and the
results of the phenomenological interactions. The $5/2^-$ state with
dominant $1p-1h$ $(p_{3/2})^{-1}(f_{5/2})^1$ configuration above the
ground state is in our calculation the $5/2^-_1$ state, while in the
phenomenological interactions it is the $5/2^-_2$ state, lying $1\mev$
higher for KB3G and $2\mev$ higher for GXPF1A.  The reason for the
difference when using 3N forces is related to the low $5/2^-$ state in
$^{49}$Ca, originating from the small effective $p_{1/2}-f_{5/2}$ gap
in our MBPT approach. Note that this effective gap is also
significantly different between the phenomenological interactions. In
turn, the $5/2^-_2$ state in the MBPT calculations has a
$(p_{3/2})^{2}_{J=2}(p_{1/2})^1$ dominant structure (on top of
$^{48}$Ca) and is therefore related to the $2^+_1$ state in $^{50}$Ca.
In all calculations it agrees with the tentatively assigned
experimental state at $2.4\mev$.  Ultimately, improved gamma-ray
spectroscopy is needed.

\subsubsection{$^{52}$Ca}

\begin{figure}
\begin{center}
\includegraphics[width=0.95\columnwidth,clip=]{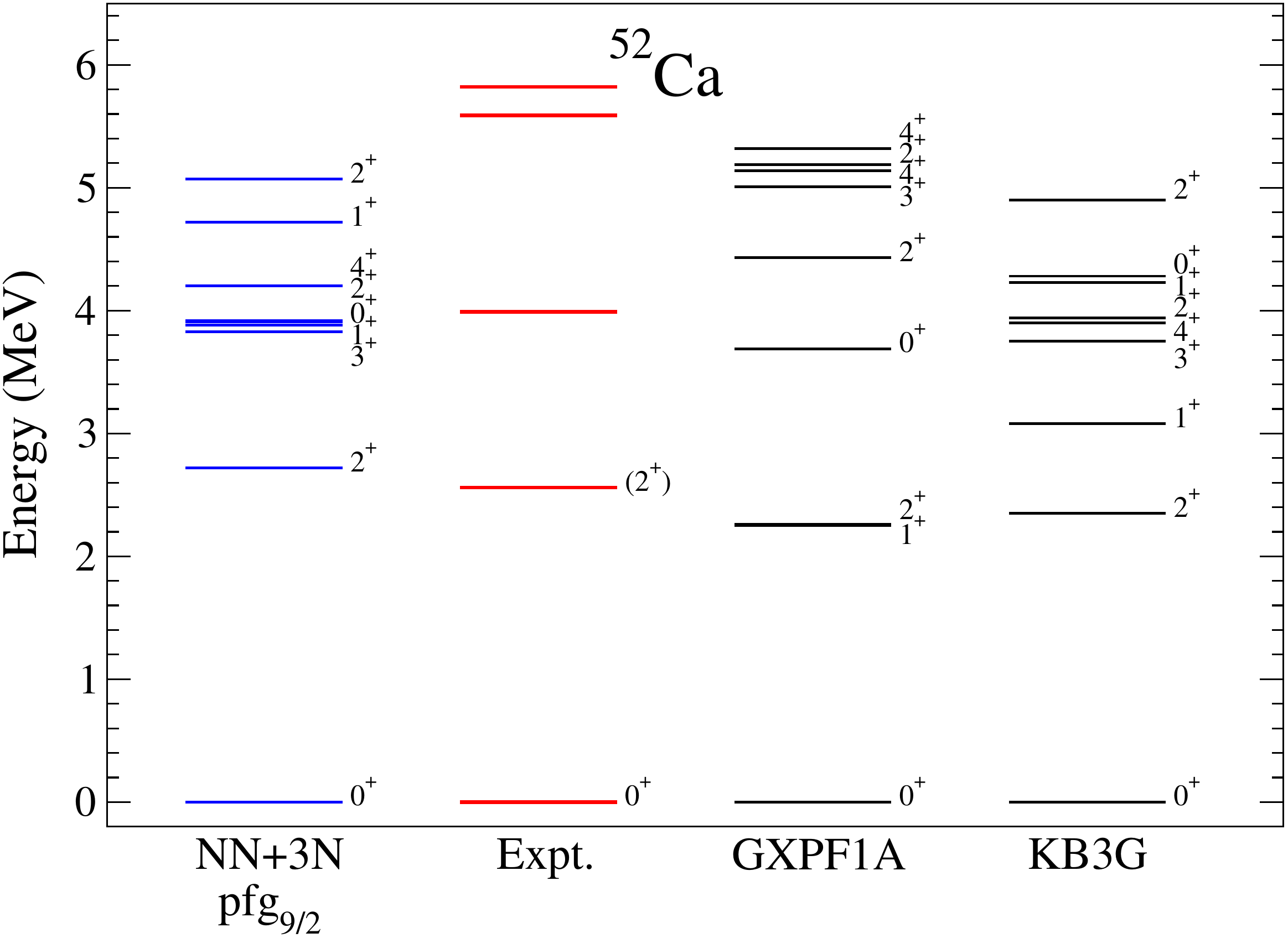}
\end{center}
\vspace{-2mm}
\caption{(Color online) Excitation energies of bound excited states in $^{52}$Ca
compared with experiment~\cite{Fornal,Rejmund} and phenomenological
interactions (labels as in Fig.~\ref{47}).\label{52}}
\end{figure}

For $^{52}$Ca, there are no spin assignments except for the ground and
first-excited state, where the large spacing was first identified as a
signature of the $N=32$ subshell closure~\cite{Huck}. The strong
$N=32$ shell closure has been unambiguously established with mass
measurements out to $^{54}$Ca, leading to a steep decrease of the
two-neutron separation energy after $^{52}$Ca~\cite{Wienholtz}. In
Fig.~\ref{52} our NN+3N calculations are compared to the 
phenomenological interactions. All agree well with the limited experimental 
data.

One striking difference between models, however, is the location of
the $1^+_1$ state, which is found in our MBPT calculations $1\mev$ and
$2\mev$ above the KB3G and GXPF1A calculations, respectively, and
hence an accurate experimental measurement would be highly
valuable. The $3^+_1$ state is also predicted quite differently
depending on the calculation; our MBPT value is in good agreement with
KB3G but more than $1\mev$ below that of GXPF1A.

\subsubsection{$^{53}$Ca}

\begin{figure}
\begin{center}
\includegraphics[width=0.95\columnwidth,clip=]{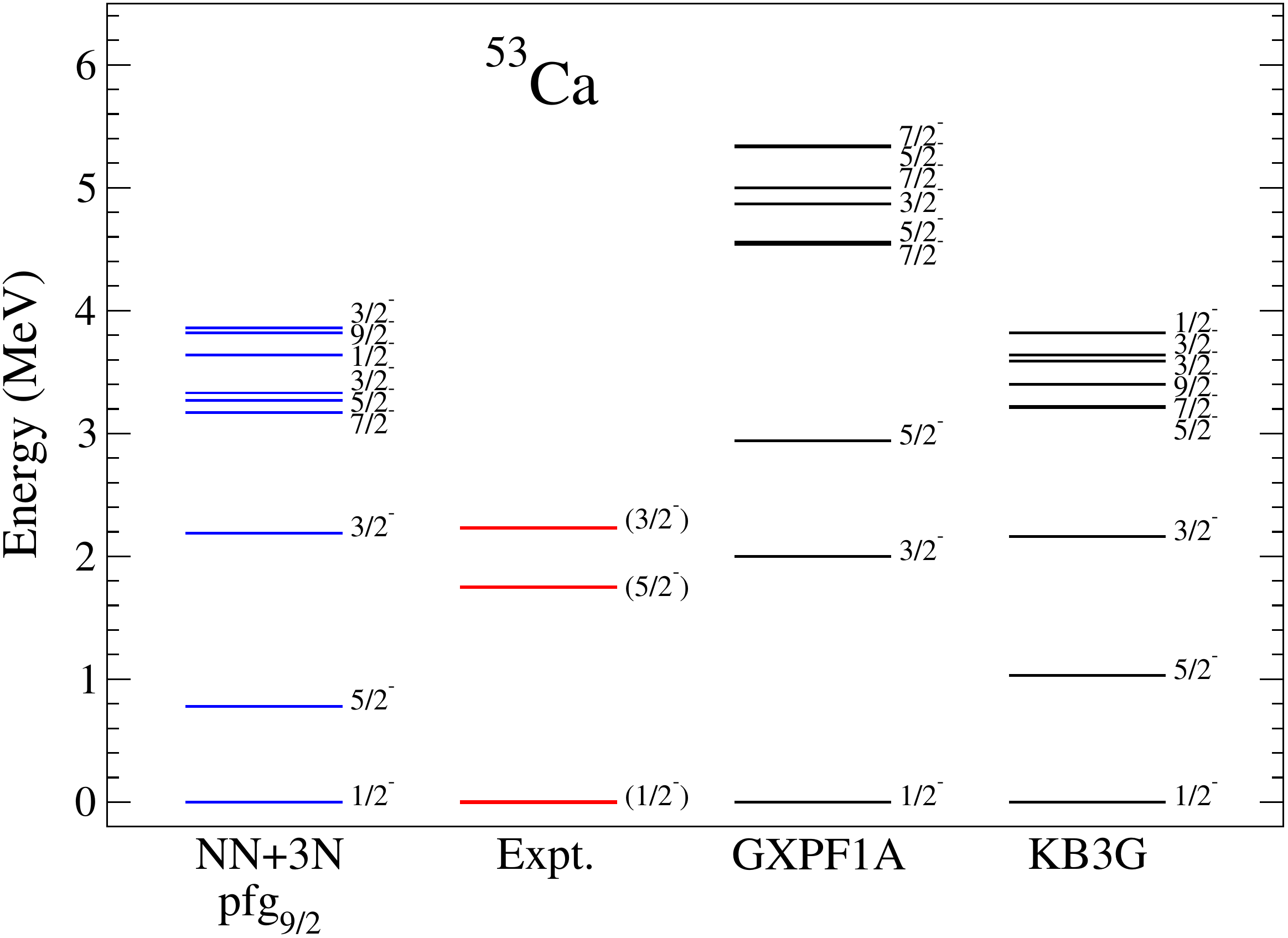}
\end{center}
\vspace{-2mm}
\caption{(Color online) Excitation energies of bound excited states in $^{53}$Ca
compared with experiment~\cite{Steppenbeck} and phenomenological 
interactions (labels as in Fig.~\ref{47}).\label{53}}
\end{figure}

Only the ground-state spin of $^{53}$Ca and the position of two
excited states are known experimentally, one of them only measured
very recently at RIKEN~\cite{Steppenbeck}.  Figure \ref{53} shows our
NN+3N calculations in the $\pfg$ space compared to the phenomenological
interactions. In this spectrum the ground state is dominated by a
$p_{1/2}$ hole in the $N=34$ closed subshell. Therefore, the
difference between the ground and first $5/2^-_1$ and $3/2^-_1$ states
will be related to the effective $p_{1/2}-f_{5/2}$ and
$p_{1/2}-p_{3/2}$ gaps and hence the strengths of the $N=32$ and
$N=34$ subshell closures, respectively. All three calculations predict
a consistent location for the lowest $3/2^-_1$ state, also in
agreement with the unassigned experimental state at $2.2 \mev$, which,
assuming this is the correct spin assignment, reflects the predictions
of the $2^+_1$ state in $^{52}$Ca.

Interestingly, the $5/2^-_1$ state appears at different excitation
energies in all calculations: $0.8\mev$ in MBPT, $1\mev$ with 
KB3G, and $3\mev$ with GXPF1A, in comparison with the state with unassigned
spin at $1.75\mev$. This shows that phenomenological interactions,
which give similar results close to stability, can extrapolate to very
different results for neutron-rich systems. In this case, the
difference is related to the small $p_{1/2}-f_{5/2}$ gap (weak $N=34$
subsell closure) predicted by KB3G, and also preferred by our MBPT
approach, in contrast with the large gap (strong $N=34$ subsell
closure) given by GXPF1A. Improved versions of GXPF1A that adjust the
$p_{1/2}^2$ and $p_{1/2}-f_{5/2}$ $T=1$ monopole matrix elements
according to the most recent experimental data, GXPF1B and
modifications, have recently become available~\cite{GXPF1B,Steppenbeck}.
They reduce the $p_{1/2}-f_{5/2}$ gap in agreement with experiment,
and predict the $5/2^-_1$ state at $1.9\mev$. The two excited states
are also in good agreement with recent CC calculations with
phenomenological 3N forces, which predict the $5/2^-_1$ and $3/2^-_1$
states at $1.9$ and $2.5\mev$, respectively~\cite{CCCa}.

\subsubsection{$^{54}$Ca}

\begin{figure}
\begin{center}
\includegraphics[width=0.95\columnwidth,clip=]{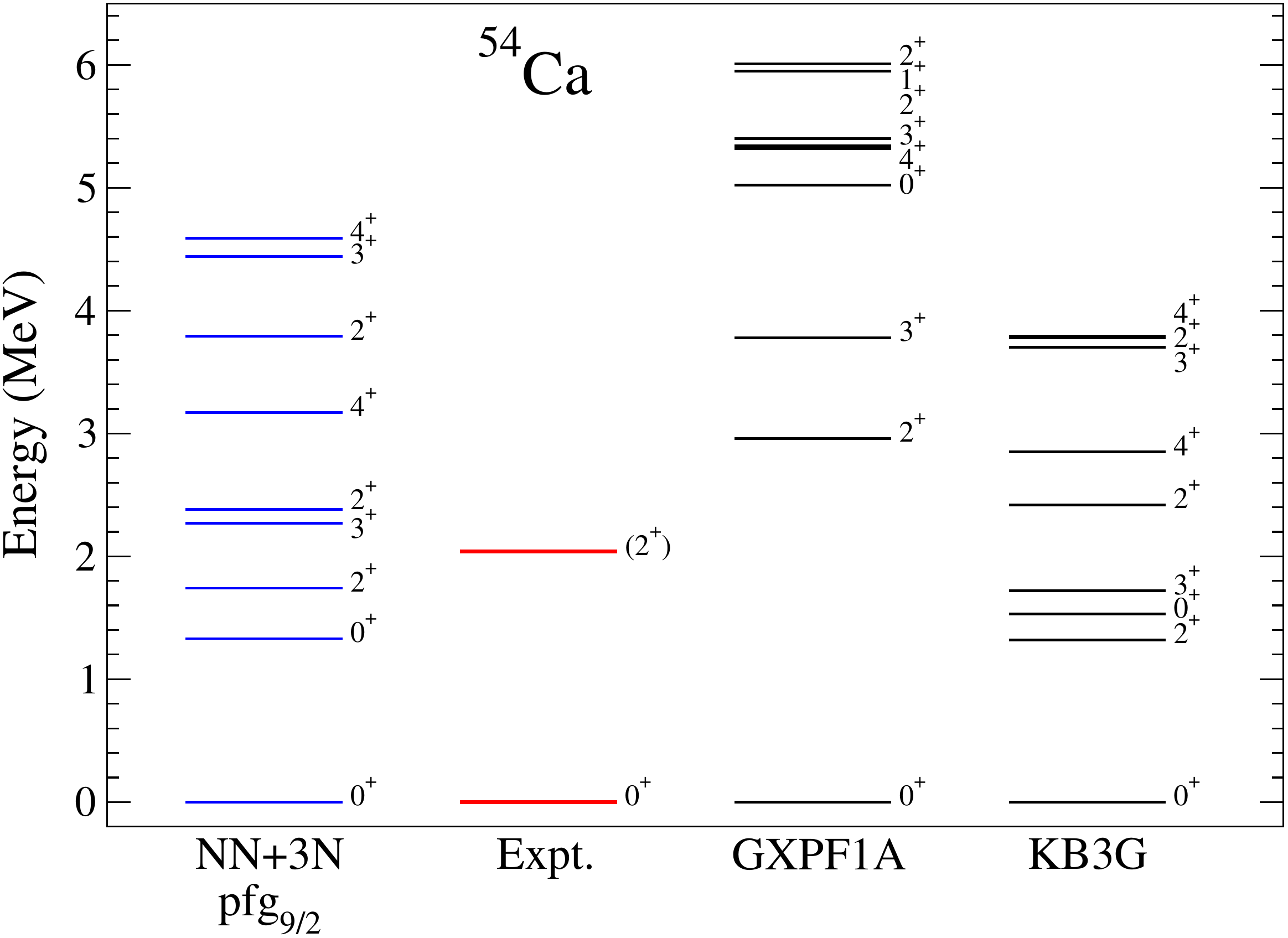}
\end{center}
\vspace{-2mm}
\caption{(Color online) Excitation energies of bound excited states in $^{54}$Ca 
compared with experiment~\cite{Steppenbeck} and phenomenological
interactions (labels as in Fig.~\ref{47}).\label{54}}
\end{figure}

\begin{figure}
\begin{center}
\includegraphics[width=0.95\columnwidth,clip=]{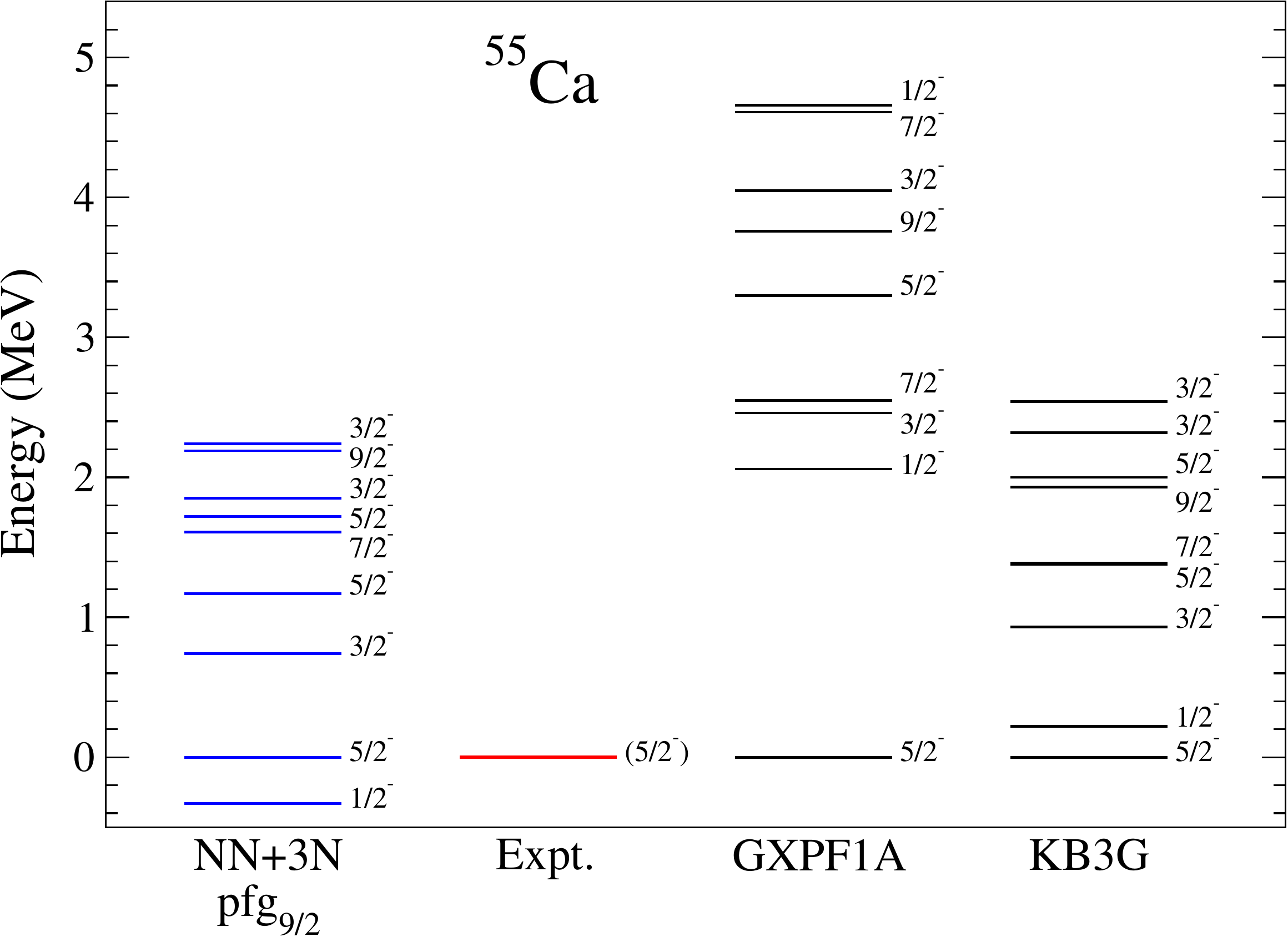}
\end{center}
\vspace{-2mm}
\caption{(Color online) Excitation energies in $^{55}$Ca 
compared with phenomenological interactions (labels as in 
Fig.~\ref{47}).\label{55}}
\end{figure}

$^{54}$Ca is the last calcium isotope for which spectroscopic data
exists. In Fig.~\ref{54}, we show our NN+3N calculations compared with
the phenomenological interactions and the recent breakthrough $2^+_1$
measurement at $2.043(19) \mev$~\cite{Steppenbeck}. Our MBPT calculations
predict several low-lying excited states, implying only a weak $N=34$
subshell closure, consistent with the spectrum in $^{53}$Ca. In
particular, the $2^+_1$ state is predicted at
$1.7\mev$~\cite{pairing}, only $300\kev$ below experiment. The $2^+_1$
excitation energy is also in very good agreement with $1.9\mev$
predicted by CC calculations with phenomenological 3N
forces~\cite{CCCa}.

The striking difference between KB3G and GXPF1A in this region is
clearly manifested in $^{54}$Ca. The recently measured $2^+_1$ energy
lies $1\mev$ below GXPF1A and $0.7\mev$ above the KB3G prediction.
The difference between these calculations is consistent with the
spectra presented in the discussion of $^{53}$Ca. As in $^{53}$Ca,
this is improved when considering the modified GXPF1B interaction,
which reproduces experiment~\cite{Steppenbeck}.

\subsubsection{$^{55}$Ca and $^{56}$Ca}

\begin{figure}
\begin{center}
\includegraphics[width=0.95\columnwidth,clip=]{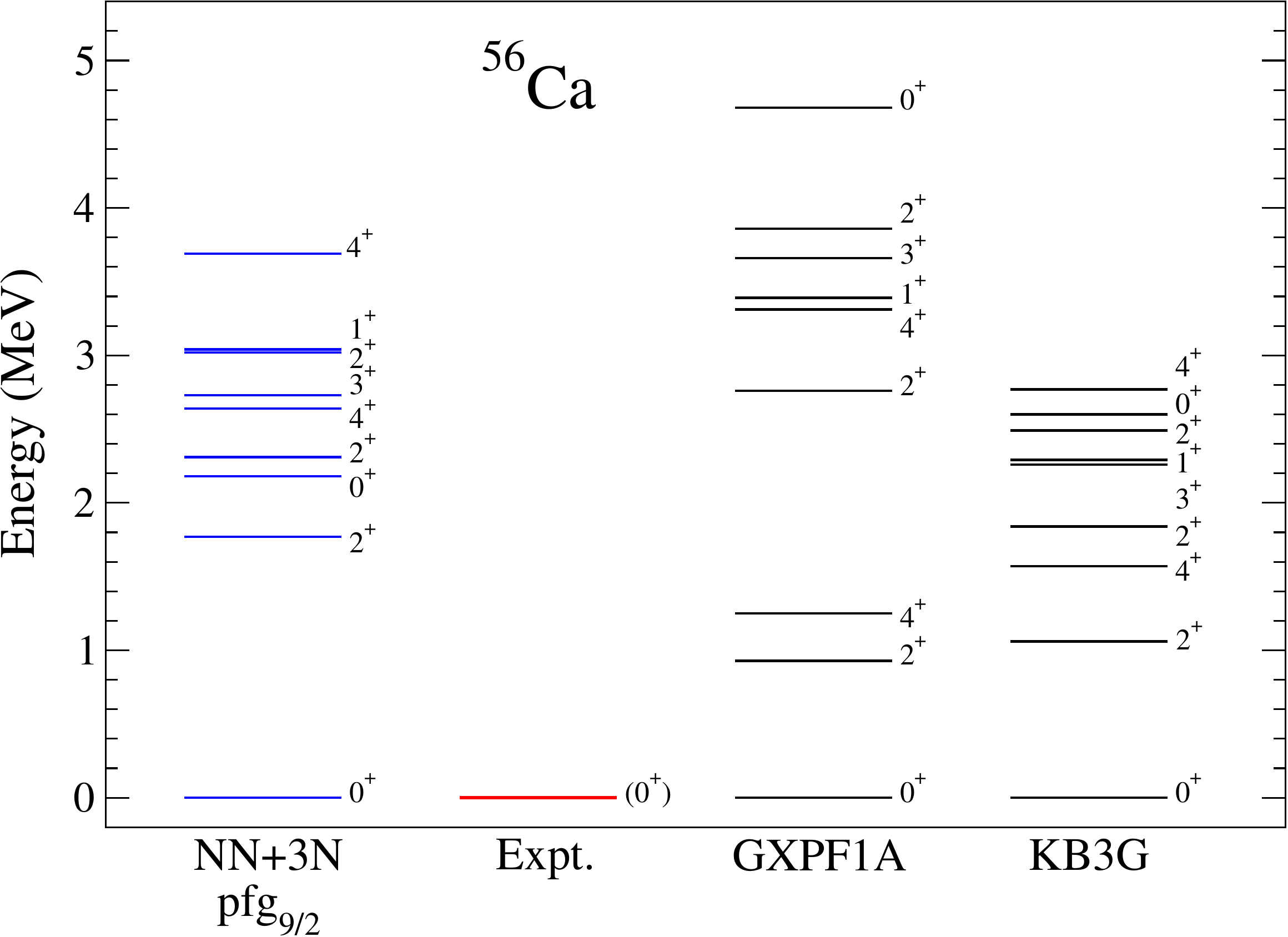}
\end{center}
\vspace{-2mm}
\caption{(Color online) Excitation energies in $^{56}$Ca 
compared with phenomenological interactions (labels as in 
Fig.~\ref{47}).\label{56}}
\end{figure}

Finally, in Figs.~\ref{55} and \ref{56} we show NN+3N predictions for the
spectra in $^{55}$Ca and $^{56}$Ca. In this region the importance of the
neutron $1d_{5/2}$ and $2s_{1/2}$ orbitals, currently not included in
our calculations, has been emphasized in Refs.~\cite{Lenzi,CCCa}. We
plan to extend our approach to include these orbitals in the valence
space and provide an improved description of very neutron-rich nuclei
near $N=40$.

Our NN+3N prediction for $^{55}$Ca in Fig.~\ref{55} largely agrees with the
predictions of the KB3G interaction, with a number of close-lying states
below 2.5$\mev$, in contrast to GXPF1A, where the first excited state
lies above 2$\mev$. However, we find a $1/2^-$ ground
state, rather than $5/2^-$ as predicted in KB3G, GXPF1A and the CC
calculations of Ref.~\cite{CCCa}. This is consistent with the picture of
a weak $N=34$ shell closure in $^{54}$Ca, where states in $^{55}$Ca are
not dominated by single-particle character.  

In $^{56}$Ca the NN+3N $2^+_1$ state is around $800\kev$ higher than in
KB3G, GXPF1A and CC~\cite{CCCa}, the same trend of higher $2^+_1$ states
than in phenomenological interactions for non-closed even-even calcium 
isotopes.  Otherwise the spacing of excited states in our NN+3N calculation is
closer to KB3G.  The differences between all three interactions in
Fig.~\ref{56} highlight the importance of experimental spectroscopic
studies beyond $^{54}$Ca.

\subsection{Electromagnetic transition strengths}

\subsubsection{$B(E2)$}

Next we study electromagnetic quadrupole ($E2$) transitions. The
results of our theoretical calculations are compared with experiment
\cite{Montanari,Raman} in Table~\ref{BE2} and with those of the phenomenological
interactions. Standard effective charges $\delta q=0.5e$, which take
into account the reduced valence space of the calculations, are
employed in all cases. Work is in progress on calculating the
consistent effective one-body operators within MBPT.

\begin{table}
\begin{center}
\begin{tabular*}{0.98\columnwidth}{@{\extracolsep{\fill}}ccccc}
\hline\hline
Transition & KB3G & GXPF1A & MBPT & Exp. \\ \hline
$^{46}$Ca: $2^+\rightarrow0^+$     & 9.2 & 9.2 & 13.3 & 25.4$\pm$4.5 \\
                                  & & & & 36.4$\pm$2.6 \\
$^{46}$Ca: $4^+\rightarrow2^+$     & 7.5 & 7.1 & 9.9 & 8.6$\pm$2.1 \\
$^{46}$Ca: $6^+\rightarrow4^+$     & 3.6 & 3.6 & 4.8 & 5.38$\pm$0.29 \\
$^{47}$Ca: $3/2^-\rightarrow7/2^-$ & 0.84& 3.6 & 1.0 & 4.0$\pm$0.2 \\
$^{48}$Ca: $2^+\rightarrow0^+$     & 11.5 & 11.9 & 10.3 & 19$\pm$6.4 \\
$^{49}$Ca: $7/2^-\rightarrow3/2^-$ & 0.41 & 4.0 & 0.22 & 0.53$\pm$0.21 \\
$^{50}$Ca: $2^+\rightarrow0^+$     & 8.9 & 9.1 & 11.2 & 7.4$\pm$0.2 \\
\hline\hline
\end{tabular*}
\end{center}
\vspace{-2mm}
\caption{Electric quadrupole transition rates in the calcium isotopes
compared with experiment~\cite{Montanari,Raman}. The $B(E2)$ values
are calculated in the $\pfg$ space with NN+3N interactions. Effective
charges $\delta q=0.5e$ are used in all calculations, which employ a
harmonic-oscillator length $b=(\hbar/m\omega)^{1/2}$ with $\hbar\omega
= (45 A^{-1/3} - 25 A^{-2/3})\mev$ and nucleon mass $m$. The
units are $e^2$\,fm$^4$.\label{BE2}}
\end{table}

Overall, we find good agreement between our valence-space calculations
and experiment. The measured $B(E2)$ transitions in Table~\ref{BE2}
vary widely (within a factor of 50), therefore obtaining agreement
within experimental error bars is particularly challenging. Indeed the
calculated $B(E2)$ values can deviate from experiment by up to a
factor of two (or a factor of four for the small
$3/2^-\rightarrow7/2^-$ transition in $^{47}$Ca), which is similar to
the phenomenological interactions.

In transitions involving $^{46}$Ca, the MBPT $B(E2)$ results are
systematically $30\%$ larger than with the phenomenological
interactions and lie closer to experiment.  Nevertheless the predicted
$2^+\rightarrow0^+$ transition in $^{46}$Ca is still half the
experimental value, pointing to important missing contributions (we
also note that the two different experimental values do not
overlap). In transitions involving $^{47}$Ca, $^{48}$Ca, and
$^{49}$Ca, the MBPT $B(E2)$ values are similar to those obtained with
KB3G, but all calculations underestimate experiment. This also applies
to GXPF1A for the $2^+\rightarrow0^+$ transition in $^{48}$Ca. For the
$3/2^-\rightarrow7/2^-$ transition in $^{47}$Ca, GXPF1A finds very
good agreement with data, with the caveat that the
$7/2^-\rightarrow3/2^-$ transition in $^{49}$Ca is overpredicted.  The
MBPT $B(E2)$ values for the $2^+\rightarrow0^+$ transition in
$^{50}$Ca are $20\%$ larger than with the phenomenological
interactions, in this case $50\%$ higher than experiment.

Finally, we emphasize that electroweak two-body currents
(meson-exchange currents) have not been included in our calculations
of electromagnetic transitions. These have been derived consistently
in chiral EFT and shown to be important for magnetic moments and
electromagnetic transitions in light nuclei~\cite{Pastore}. For
axial-vector weak interactions, chiral two-body currents have been
applied to medium-mass nuclei, showing that they provide important
contributions to the quenching of Gamow-Teller
transitions~\cite{2bcGT}. Work is in progress to extend this to
electromagnetic currents and to calculate effective operators
consistently in MBPT.

\subsubsection{$B(M1)$}

\begin{figure}
\begin{center}
\includegraphics[width=\columnwidth,clip=]{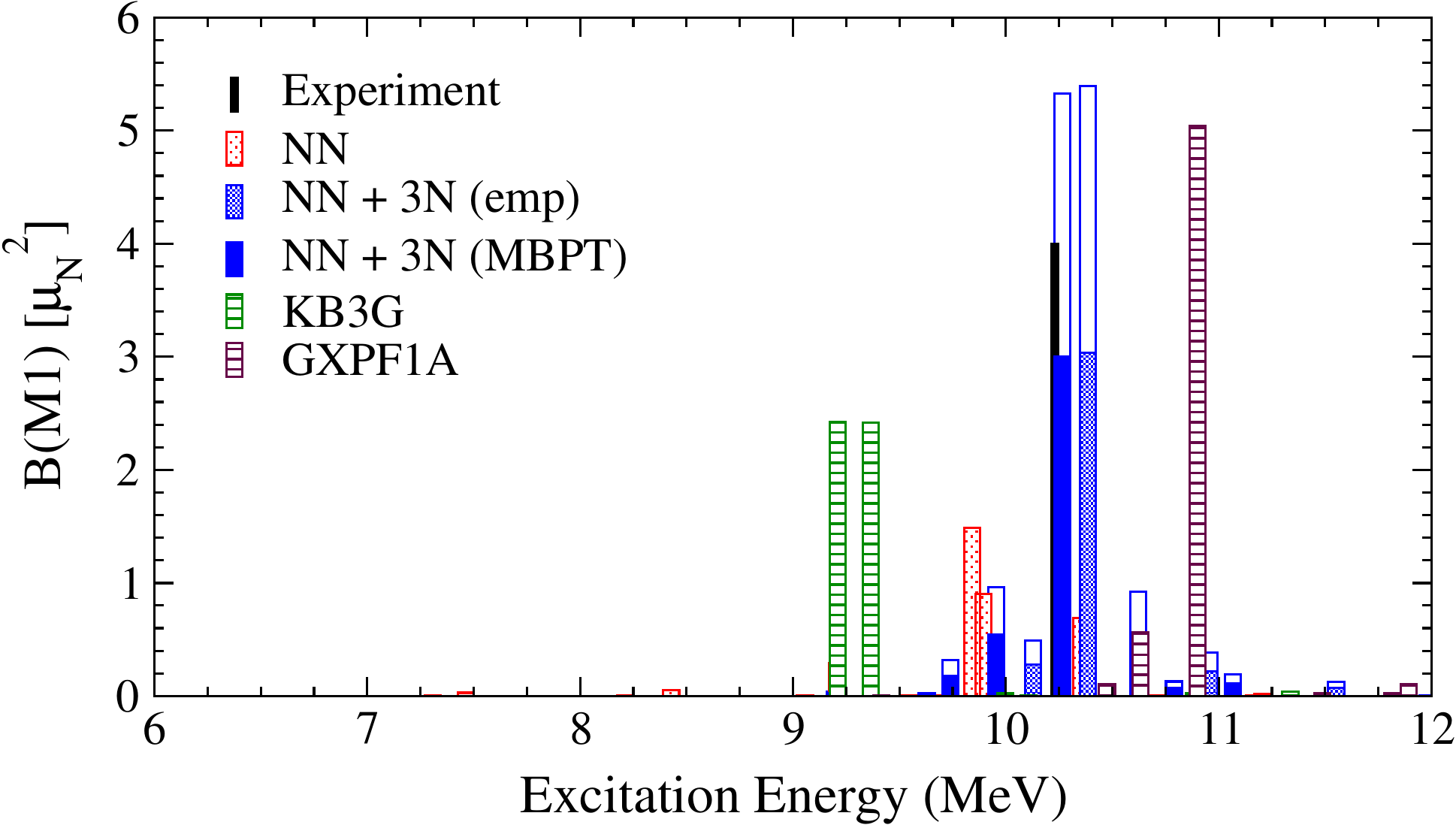}
\end{center}
\vspace{-2mm}
\caption{(Color online) Magnetic dipole transition rates from the ground state to 
$1^+$ excited states in $^{48}$Ca compared with experiment~\cite{48CaM1}.
The $B(M1)$ values are calculated in the $\pfg$ space with NN+3N
interactions. Spin $g$ factors are quenched by 0.75, except for the
empty blue bars.\label{M1}}
\end{figure}

The magnetic dipole ($M1$) transition between the ground state of
$^{48}$Ca and $1^+$ excited states is compared to experiment \cite{48CaM1}
in Fig.~\ref{M1}.  All results use spin $g$ factors with empirical
quenching $q=0.75$, except for the empty blue bars, which give the
NN+3N results without quenching. The main conclusions of previous work
using 3N forces to first order in MBPT still apply~\cite{Calcium}. With NN 
forces only, the $B(M1)$ strength is strongly fragmented with a central value 
well below experiment.  For the phenomenological interactions, GXPF1A 
finds a concentrated peak, while for KB3G the total strength is evenly 
fragmented between two peaks. We note, however, that the related KB3 
interaction \cite{kb3} finds no fragmentation, which highlights the particular sensitivity 
of this transition to the valence-space Hamiltonian. The peaks
lie about 1 $\mev$ lower than experiment for KB3G, and around 700 $\kev$
higher for GXPF1A.  When 3N 
forces are included in the $\pfg$ space, agreement with experimental data
is clearly improved, with a concentrated peak very close to the
experimental value. The effects of the improved treatment of 3N forces
over that in Ref.~\cite{Calcium} is an increase of the excitation
energy of the peak transition by $\sim 500\kev$. The degree of
single-particle character of the $M1$ transition, once 3N forces are
included, is similar for calculated (MBPT) or empirical (emp)
SPEs. Finally, it is interesting that the NN+3N calculations do not
require a strong quenching of spin $g$ factors. While the unquenched
NN+3N results indeed overpredict the $M1$ transition strength, they
are already at the level of the quenched GXPF1A value; only a modest
quenching of 0.9 would be needed to reproduce the experimental strength.

\subsection{Residual 3N forces}
\label{residual3N}

\begin{figure}
\begin{center}
\includegraphics[width=0.45\textwidth,clip=]{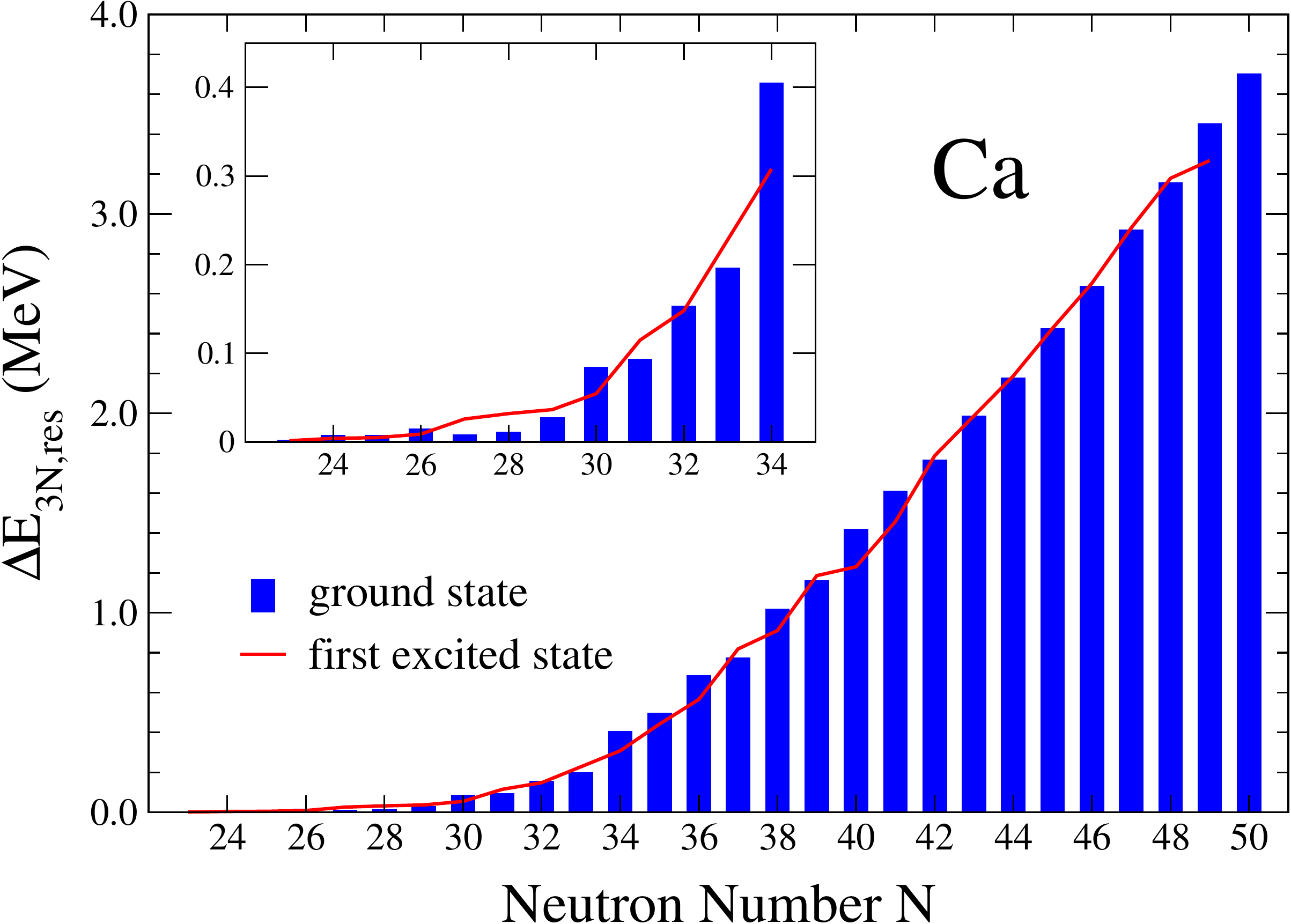}
\end{center}
\vspace{-2mm}
\caption{(Color online) First-order contribution from residual 3N forces to the 
energies of the ground state (bars) and first excited state
(solid line) of the calcium isotopes in the extended $\pfg$ valence
space. \label{residual}}
\end{figure}

We have also explored residual 3N forces, which arise between three
valence nucleons in addition to their one- and two-body normal-ordered
contributions. Note that in the shell model the normal ordering is
performed with respect to the $^{40}$Ca core. This does not take into
account the contributions from MBPT outside of the valence space. In
general, residual 3N forces are expected to lead to small corrections
due to phase-space considerations for normal Fermi
systems~\cite{Fermi}, because their effects are suppressed by the
number of valence particles to the number of particles in the core. We
have calculated the contributions from residual 3N forces at first
order in perturbation theory, using the states obtained from NN+3N
interactions at the normal-ordered one- and two-body level (as
discussed in the previous sections).

In Fig.~\ref{residual} we show the resulting energies $\Delta 
E_{\rm 3N,res} = \la \psi_n| V^{\rm 3N}_{\rm res}|\psi_n \ra$ for the ground
state and first excited state calculated in the extended
$\pfg$ space.  As expected from neutron-matter
calculations~\cite{nm,EPJAreview}, residual 3N forces are
repulsive. Their role is very minor for the lighter isotopes, but
their contribution is amplified with neutron number, increasing to a
maximum contribution of $3.7\mev$ in $^{70}$Ca. For the mid shell
$A=55-58$ isotopes, only a subset of residual 3N matrix elements was
used. From tests in other isotopes, we estimate the difference to
using the full set to be less than $5\%$ of the total residual 3N
energy contribution.

We compare the effects of residual 3N forces with the NN+3N
ground-state energies of Fig.~\ref{gs} at the normal-ordered two-body
level. This shows that residual 3N forces provide small corrections to
the ground-state energy, ranging from at most $3\%$ (in $^{56}$Ca) for
the isotopes discussed in Sect.~\ref{spectra} to $9\%$ in
$^{70}$Ca. This justifies our perturbative estimate of residual 3N
forces. As shown in Fig.~\ref{residual}, residual 3N contributions are
similar for excited states as for ground states. Therefore, residual
3N forces lead to even smaller corrections for excitation energies, of
$20\kev$ for $^{48}$Ca, increasing to $100\kev$ in $^{54}$Ca, and
never exceeding $200\kev$. As a result, the uncertainty from not
including residual 3N forces is small for the results presented in the
previous sections.

\section{Summary}

We have presented a comprehensive study of excited-state properties of
calcium isotopes based on chiral NN+3N interactions. The theoretical
approach has been discussed in detail, focusing on convergence
properties in the MBPT framework, benchmarking against ab initio CC
theory for NN interactions, and exploring the role of residual 3N
forces. We have presented results for ground-state energies and
spectra for neutron-rich isotopes to $^{56}$Ca, where 3N forces were
shown to be key to understand the experimental structures. With both
3N forces and an extended $\pfg$ valence space, we obtain a good level
of agreement with experiment, where the extended space is especially
important for $N \geqslant 28$. We have also studied electromagnetic
$E2$ and $M1$ transitions, finding that experimental data are well
described by our calculations. Where data does not exist, our results
provide predictions for unexplored properties of neutron-rich calcium
isotopes.

Future work will include studies of the theoretical uncertainties due
to the input Hamiltonian and the RG or SRG evolution.  In
addition, the recent development of the IMSRG for open
shell~\cite{IMSRGopen,IMSRGSM} and CC calculations of effective
interactions~\cite{CCSM} enable nonperturbative derivations of
valence-space Hamiltonians, which will also provide benchmarks for MBPT
calculations.

\begin{acknowledgments}
We thank G.\ Hagen for helpful discussions.  This work was supported
by the BMBF Contract No.~06DA70471, the DFG through Grant SFB~634, the
ERC Grant No.~307986 STRONGINT, and the Helmholtz Alliance HA216/EMMI.
Computations were performed on JUROPA at the J\"ulich Supercomputing Center 
and on Kraken at the National Institute for Computational Sciences.
\end{acknowledgments}

\appendix

\section{Spectra of light calcium isotopes $A<47$}
\label{light_spectra}

\begin{figure}
\begin{center}
\includegraphics[width=0.9\columnwidth,clip=]{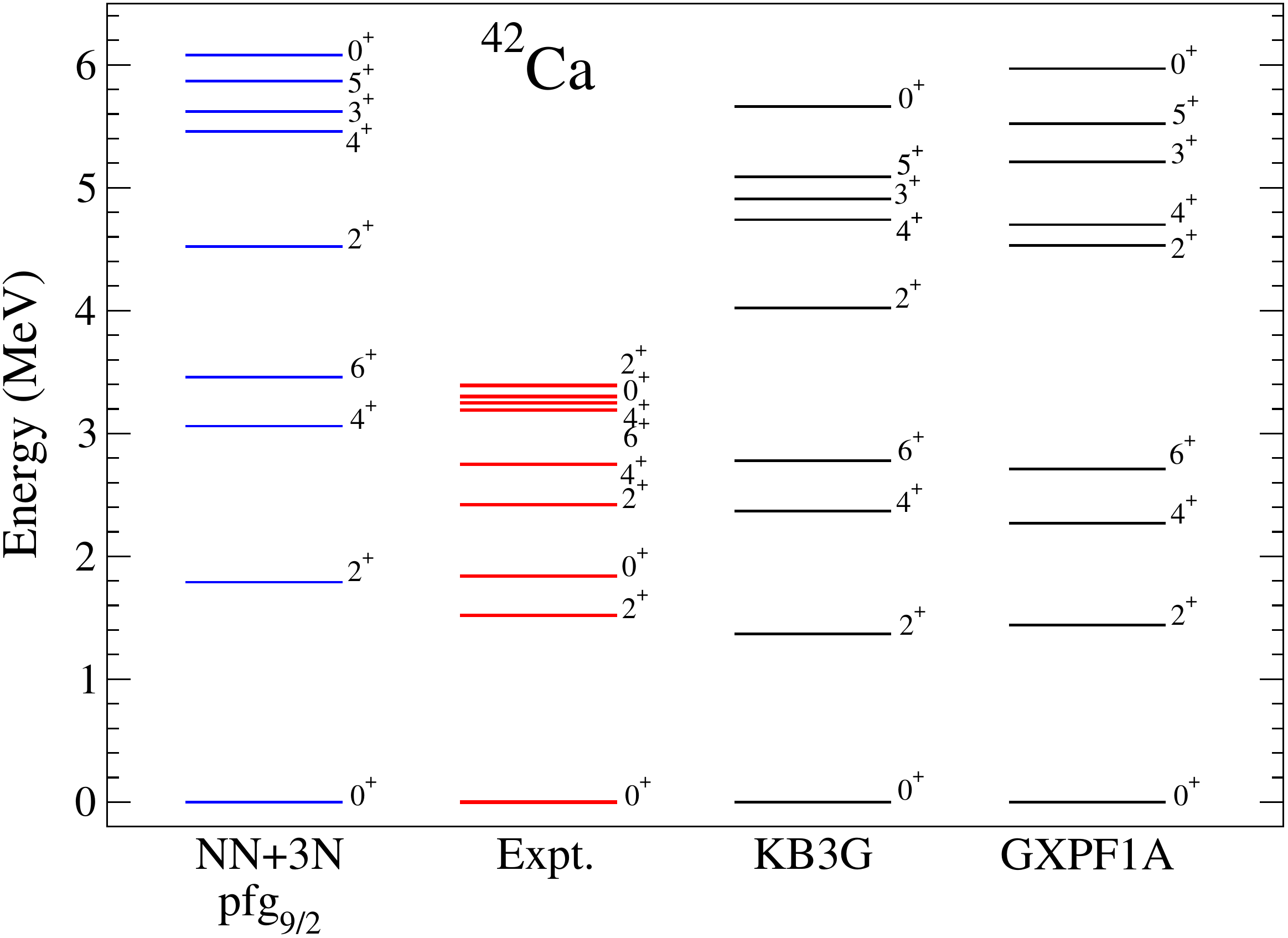}
\end{center}
\vspace{-2mm}
\caption{(Color online) Excitation energies of bound excited states in $^{42}$Ca
compared with experiment and phenomenological interactions (labels
as in Fig.~\ref{47}).\label{42}}
\end{figure}

\begin{figure}
\begin{center}
\includegraphics[width=0.9\columnwidth,clip=]{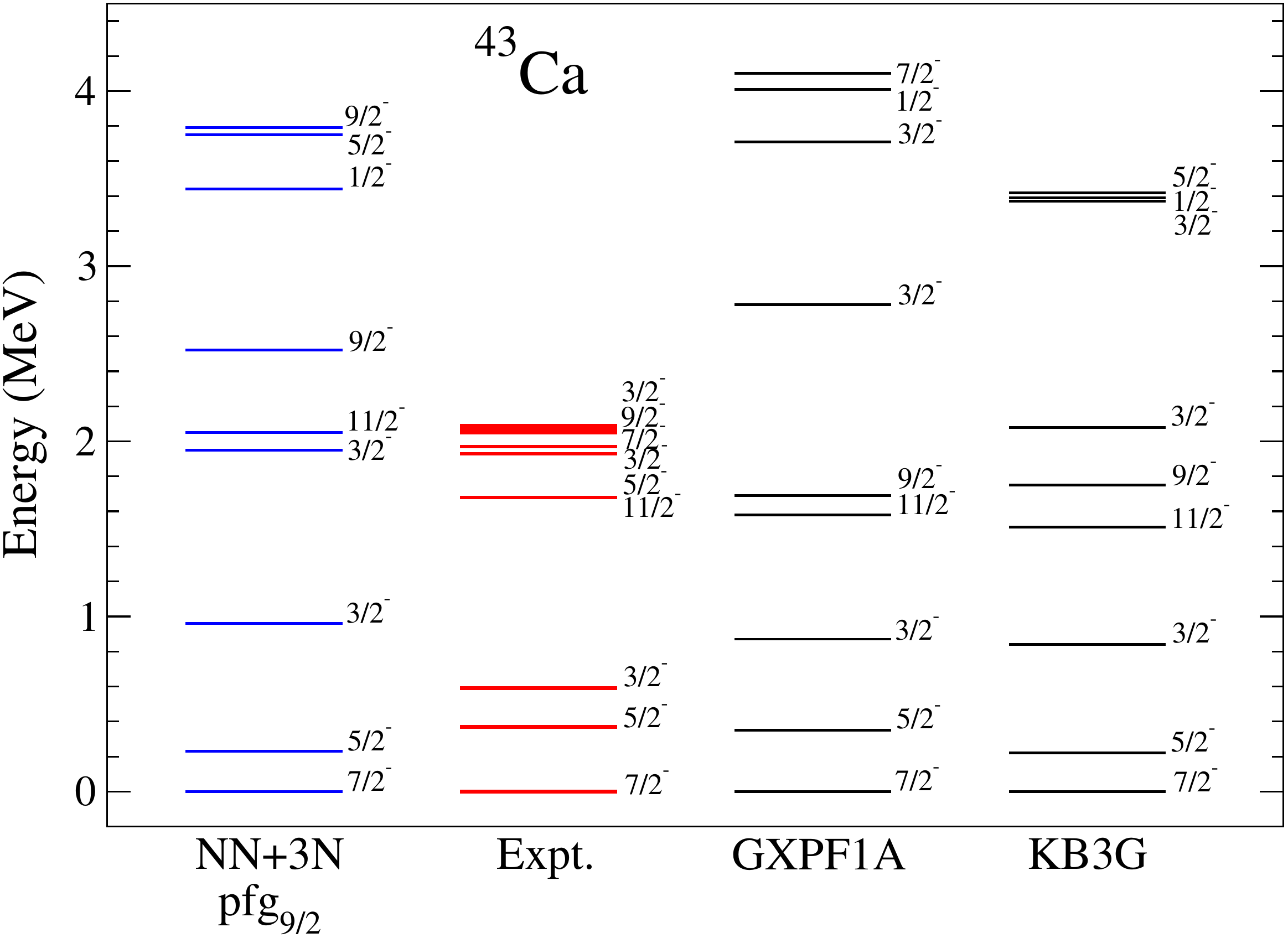}
\end{center}
\vspace{-2mm}
\caption{(Color online) Excitation energies of bound excited states in $^{43}$Ca
compared with experiment and phenomenological interactions (labels
as in Fig.~\ref{47}).\label{43}}
\end{figure}

\begin{figure}
\begin{center}
\includegraphics[width=0.9\columnwidth,clip=]{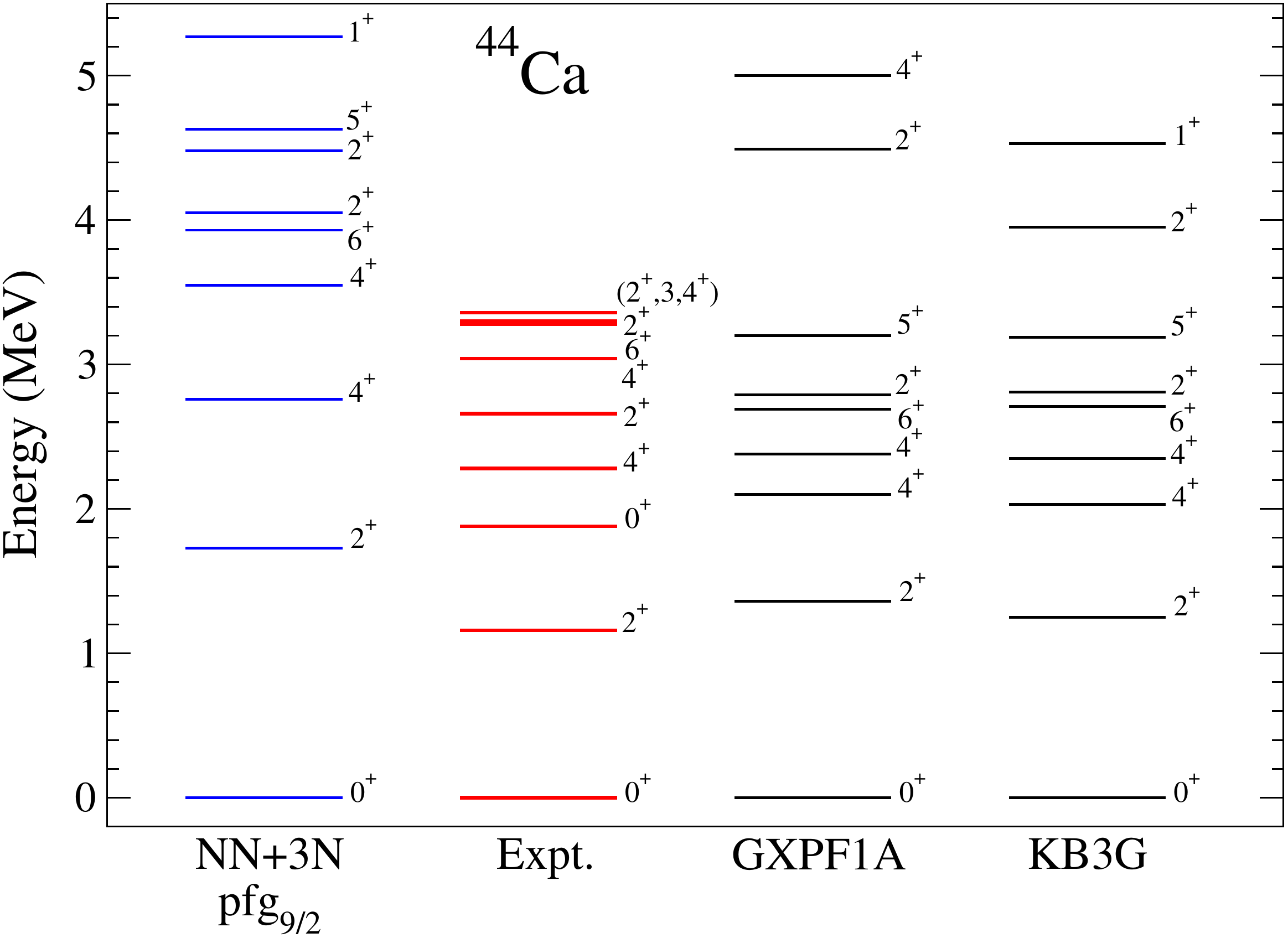}
\end{center}
\vspace{-2mm}
\caption{(Color online) Excitation energies of bound excited states in $^{44}$Ca
compared with experiment and phenomenological interactions (labels
as in Fig.~\ref{47}).\label{44}}
\end{figure}

\begin{figure}
\begin{center}
\includegraphics[width=0.9\columnwidth,clip=]{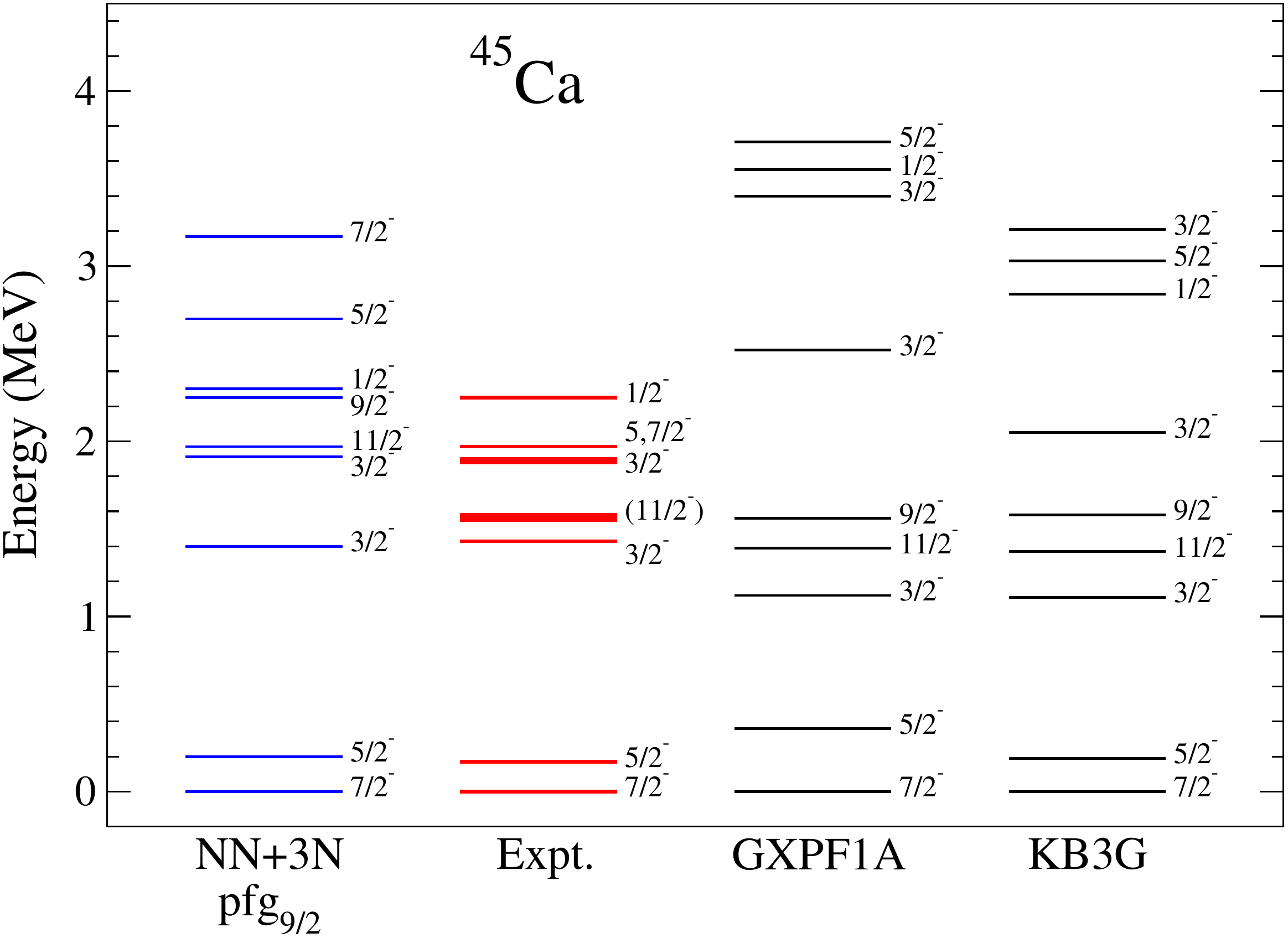}
\end{center}
\vspace{-2mm}
\caption{(Color online) Excitation energies of bound excited states in $^{45}$Ca
compared with experiment and phenomenological interactions (labels
as in Fig.~\ref{47}).\label{45}}
\end{figure}

\begin{figure}
\begin{center}
\includegraphics[width=0.9\columnwidth,clip=]{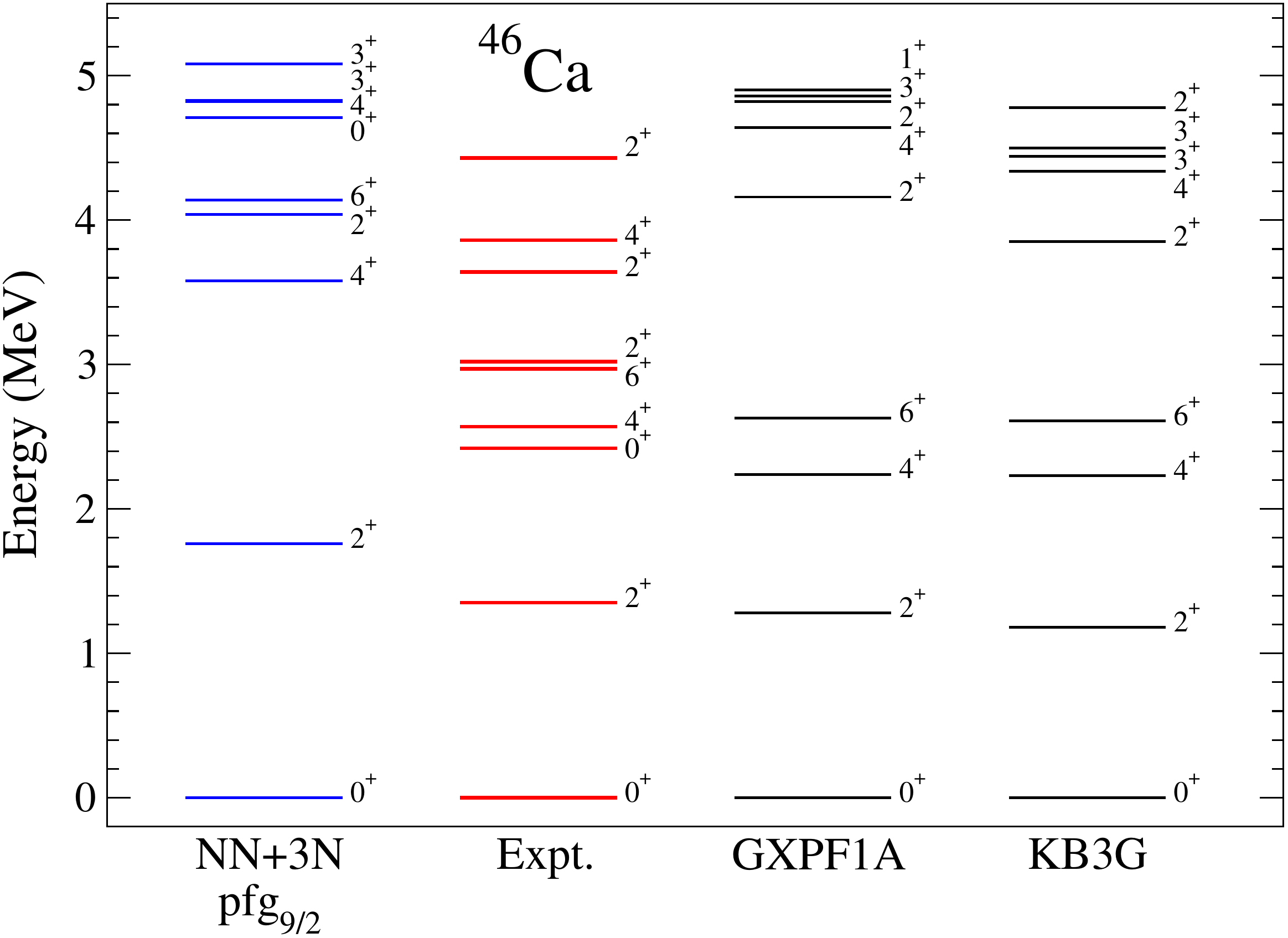}
\end{center}
\vspace{-2mm}
\caption{(Color online) Excitation energies of bound excited states in $^{46}$Ca
compared with experiment and phenomenological interactions (labels
as in Fig.~\ref{47}).\label{46}}
\end{figure}

We present the spectra for the lighter $^{42-46}$Ca in
Figs.~\ref{42}--\ref{46}.  Our results are in good agreement with
experiment~\cite{ENSDF}, and generally exhibit quality comparable
to phenomenological interactions.  The only exceptions are the 
$4^+, 6^+$ states in $^{46}$Ca, $\sim 1 \mev$ higher than in 
experiment, in contrast to phenomenology.  Note that some
excited states, like the $0^+_2$ states in $^{42,44,46}$Ca, 
are expected to be dominated by $sd$ shell degrees of freedom, and 
are therefore not present in our theoretical results.

\end{document}